\documentclass[prl, twocolumn, superscriptaddress, nofootinbib]{revtex4-2}

\usepackage{graphicx}% Include figure files
\usepackage{dcolumn}% Align table columns on decimal point
\usepackage{bm}% bold math
\usepackage{lipsum}
\usepackage{amsmath}
\usepackage{gensymb}
\usepackage{wasysym}
\usepackage{float}

\begin{document}

\preprint{APS/123-QED}

\title{Metasurface-Enabled Astronomical Polarimetry}

\author{Lisa W. Li}
\affiliation{Department of Electrical and Computer Engineering, University of California San Diego, La Jolla, CA 92093 USA}
\author{Phillip H.H. Oakley}
\affiliation{BAE Systems Inc. Space and Mission Systems, Boulder, CO 80301 USA}
\author{Rebecca N. Schindhelm}
\affiliation{BAE Systems Inc. Space and Mission Systems, Boulder, CO 80301 USA}
\author{Sean G. Sellers}
\affiliation{New Mexico State University, Las Cruces, NM 88003 USA}
\author{Roberto Casini}
\affiliation{High Altitude Observatory, Boulder, CO 80301 USA}
\author{Noah A. Rubin}\email{noahrubin@ucsd.edu}
\affiliation{Department of Electrical and Computer Engineering, University of California San Diego, La Jolla, CA 92093 USA}

\date{\today}% It is always \today, today,

\begin{abstract}
In the last decade or so, metasurface optical components have received considerable scientific and industrial interest for a variety of applications. The miniaturization afforded by metasurfaces could benefit astronomy in particular (which is an often-cited potential application area for metasurfaces). However, few developed examples in which metasurface components offer a unique benefit to astronomical instrumentation---substantiated by the production of scientific data---have been shown. Here, we present the Solar Imaging Metasurface Polarimeter (SIMPol), a first-of-its-kind telescope for snapshot imaging polarimetry of the sun around a Sr I line at 460.7 nm enabled by a metasurface polarization-analyzing grating. This high-performance grating exhibits an overall efficiency of nearly 70\% and high polarization contrast (diattenuation) across its four observation channels. We demonstrate SIMPol's integration into a major observatory telescope facility with two different imaging modes. In both cases, Zeeman polarization signatures were clearly observed in two adjacent spectral lines of Fe I and Sr I around 460.7 nm. This work demonstrates an early success of metasurface polarization optics in a real application in astronomical instrumentation (here, polarimetric observations of the solar atmosphere), and heralds the application of metasurfaces and emergent nanophotonic technologies in astronomy more broadly.
\end{abstract}

\maketitle

%\tableofcontents

\section{Introduction}
\label{sec:intro}

Metasurfaces are diffractive optical components intended to spatially modulate the phase, amplitude, and/or polarization state of light using subwavelength features~\cite{Dorrah2022}. These components are enabled by and scalably produced with modern semiconductor fabrication techniques, and have formed the basis of intense academic, scientific, and industrial research over the past decade. The ability to spatially structure light's polarization state counts among the most unique features of metasurfaces, one which most distinguishes metasurfaces from diffractive optics technologies of the past. This has led to an active research field of metasurface polarization optics, with a variety of new polarization-dependent wavefront shaping and proof-of-principle devices shown in recent years~\cite{Rubin2021a, Mueller2017, Arbabi2015}.

The primary stated advantages of metasurface-like optical components are usually in terms of size, weight, and (with polarization in particular) the ability to combine the function of what would otherwise necessitate many optical components into one flat multi-functional metasurface. Astronomical instrumentation is a natural application which stands to benefit from all of these---astronomical observatories often have some of the strictest performance and function requirements which have historically driven advances in optical technology. In the case of space-based instrumentation, concerns of size, weight, and speed of acquisition are paramount. It is unsurprising, then, that astronomy and telescopes are often cited as potential application areas for metasurface optics. However, the application of metasurfaces to astronomical instrumentation is at a nascent stage. Applications that have been demonstrated are of a very simple nature (e.g., the use of a metalens-like element to photograph the surface of the moon~\cite{Zhang2023, Park2024, Majumder2025}) and do not yet constitute scientific proofs-of-concept. Moreover, it is unclear whether sweeping attempts at miniaturization, particularly at the level of attempting to replace a telescope's objective lens itself, can ever achieve the required level of stray light, aberration, and throughput performance needed for the exacting nature of observational astronomy. Nearer-term applications (in which metasurfaces are integrated into larger, more traditional optical systems) should then seek to benefit from unique optical functions (in particular, non-imaging applications) only afforded by metasurface-like components, rather than attempting to entirely replace other mature, well-established optics.

Recent developments in metasurface polarization optics~\cite{Rubin2021a}---with no direct analogues in conventional optics or in past generations of diffractive optical elements from which astronomy might have already benefited in decades past---form a logical starting point. In particular, metasurfaces have shown utility in \emph{polarimeters}---optical instruments which measure the polarization state of light~\cite{Rubin2019, Rubin2022, Arbabi2018}. Metasurface polarization-analyzing gratings can be designed to direct incident light into a specified set of diffraction orders with each order acting in a  polarizer-like fashion, analyzing for a designer-specified polarization state. If at least four of these orders are measured, the polarization state in the form of the full Stokes vector $\vec{S}= \begin{pmatrix} I & Q & U & V \end{pmatrix}^T$ of incident light can be determined. If such a grating is placed in a pupil plane of an optical system, this analysis can be carried out over an imaging field-of-view (FOV), permitting imaging polarimetry with a single component.

Polarimetric observations are of wide interest in astronomy, for instance in the characterization of stellar atmospheres~\cite{Trippe2014}, interstellar~\cite{Kolokolova2015, Kolokolova2015a} and cometary dust~\cite{Kolokolova1997}, and recently for the potential characterization of exoplanets and their atmospheres~\cite{Rossi2017} (including for chiral bio-signatures~\cite{Sparks2021}). However, it is in solar astronomy where polarimetry is most prominent and where it serves as a core observational technique of significant importance. This is owed to the magnetic fields present on the Sun and the crucial (and ill-understood) role these play in the formation and evolution of magnetically active regions, flares, and coronal mass ejections (which are the main drivers of space weather, and can threaten electronic and communication infrastructure on Earth). Polarized light from the Sun is a remotely observable signature of the magnetic properties of its atmosphere, first-and-foremost through the well-known Zeeman effect (first observed in polarization on the Sun by Hale in 1908~\cite{Hale1979}). For high magnetic field strengths, the Zeeman effect can be resolved spectroscopically as a line splitting, but for weaker magnetic fields producing spectral splitting irresolvable beneath a spectral feature's thermally broadened profile, a full-Stokes polarization measurement is needed to fully characterize the magnetic field vector. The diagnosis of even weaker fields characterizing the ``quiet'' magnetism outside of solar active regions can also be attained from polarimetric measurements of the Hanle effect, closely related to the Zeeman effect, in which the line polarization resonance scattering is modified by the presence of a magnetic field. This is the case for the Sr~I line at 460.7\,nm chosen for SIMPol, and for many other spectral diagnostics formed within the tenuous gases of the solar chromosphere.

In the language of solar astronomy, any polarimeter requires a ``modulation scheme'' by which the polarization-analyzing optics of an instrument acquire the requisite multiple measurements needed to determine the Stokes vector $\vec{S}$~\cite{Iglesias2019}. Most commonly, solar polarimeters use \emph{temporal} modulation whereby a waveplate retarder is mechanically rotated (or, less commonly, a liquid crystal variable retarder is electrically switched)---both in ground-based and space-based instruments. 

Temporal polarization modulation schemes present two principal disadvantages: First, they often require a mechanically moving part. For space observations in particular, the use of mechanisms poses a significant mission risk. For example, the HINODE solar mission~\cite{Kosugi2008, Tsuneta2008} launched to low-Earth orbit contained four instruments sharing a common rotating half-waveplate in the optical path from the telescope. All instruments aboard would be compromised were its rotation mechanism to fail. This scenario is not merely hypothetical: In 2010, the Solar Ultraviolet Magnetograph Investigation (SUMI) sounding rocket \cite{West2011} mission---which targeted the linear and circular polarization of the Mg II h-k line pairs of the solar chromosphere at 280 nm---experienced a failure of its rotating waveplate retarder mechanism, hindering the mission's scientific goals. 

Second, a temporal modulation scheme by definition produces a final Stokes vector image formed from data frames acquired at different times. Any time-varying features between frames are embedded into the final data product. These temporal variations could be real in origin (e.g., the dynamics of the evolving magnetic fields on the Sun observed at high spectral and spatial resolution) or undesirable artifacts (e.g., atmospheric seeing in ground-based observations, or the pointing jitter of a space-borne telescope platform in orbit)\footnote{We note that these issues are often addressed in part by ``dual-beam'' polarimeter architectures in which a polarizing beam splitter is used to send two simultaneous orthogonal states of polarization along two different optical paths. The two main advantages of the dual-beam architecture are that essentially no photons are lost to a polarization analyzer, and that signal fluctuations may be tracked during the data acquisition, which would otherwise be interpreted as spurious polarimetric properties of the target. These techniques still require a mechanism and multiple data acquisitions during which the time-varying artifacts mentioned here could again compromise the data.}. The latter are always a risk for polarimetric observations as these become encoded as polarization ``cross-talk'' into the set of modulated signals, compromising the measurement of often subtle (and small in magnitude) polarization signals from the observed object. In current generation solar astronomy missions, significant resources are routinely invested to develop pointing stability control systems (fast scanning mirrors and gimbals for satellite buses) or advanced adaptive optics to address this issue. These solutions often far exceed the cost of the polarimeter optics being stabilized to begin with.

These issues are addressed by \emph{spatial} modulation schemes, in which different optical paths simultaneously implement the required polarization measurements. No active mechanisms are required in these snapshot polarimetry systems so measurements can proceed faster---what time-varying effects still exist manifest as common-mode artifacts in the data product that simultaneously affect the full set of modulated signals necessary to infer the target's Stokes vector. This entirely eliminates temporal-variation-induced sources of polarization cross-talk. However, such schemes have so far not seen widespread adoption in solar instrumentation as they generally require significant optical hardware to implement such as complex configurations of beamsplitters (especially for full Stokes measurements) and complicate instrumental calibration requirements~\cite{Iglesias2019}. The multiple (generally orthogonal) optical paths traditionally required by spatial modulation schemes place substantial burdens on the thermal and mechanical stability requirements of the instrument.

In this work, we show that metasurfaces present a solution here, enabling spatially-modulated solar polarimetry with its accordant advantages with a single component located along a common optical path. We designed the Solar Imaging Metasurface Polarimeter (SIMPol), a refractive telescope implementing snapshot full-Stokes polarimetry around the Sr I photospheric absorption line feature at 460.7 nm. This capability is enabled by a single passive component: a dielectric metasurface polarization grating (MPG) placed in the telescope's pupil plane. The MPG forms four spatially separate, polarization-analyzed images on the sensor which, when recombined pixel-wise, give a Stokes vector $\vec{S}$ image. This high-performing grating achieves nearly 70\% overall efficiency with less than 1\% of incident energy directed to the central zero order, and diattenuation on all imaged orders exceeding 98\%.  Given the possible utility of this technology for future flight missions, we additionally space-qualify this grating through environmental testing (see below and supplement).

Finally, SIMPol is integrated with the Dunn Solar Telescope of the Sunspot Solar Observatory on Sacramento Peak (Sunspot, NM, USA). Using two distinct polarimetric imaging modes for the spectral analysis of the signal---a tunable Fabry-Perot etalon and grating spectrometer---we detected and spatially resolved the Zeeman effect signatures of the two spectral lines of Sr I and Fe I in several sunspots, demonstrating the ability of this new instrument and the metasurface technology by which it is enabled to produce quantitative scientific observations. This work demonstrates an early use case in astronomical instrumentation which leverages the significant and nuanced advantages offered by metasurfaces for optical systems.

\section{SIMPol: The Solar Imaging Metasurface Polarimeter}
\label{sec:SIMPol}

\begin{figure*}[t]
    \centering
    \includegraphics[width=\textwidth]{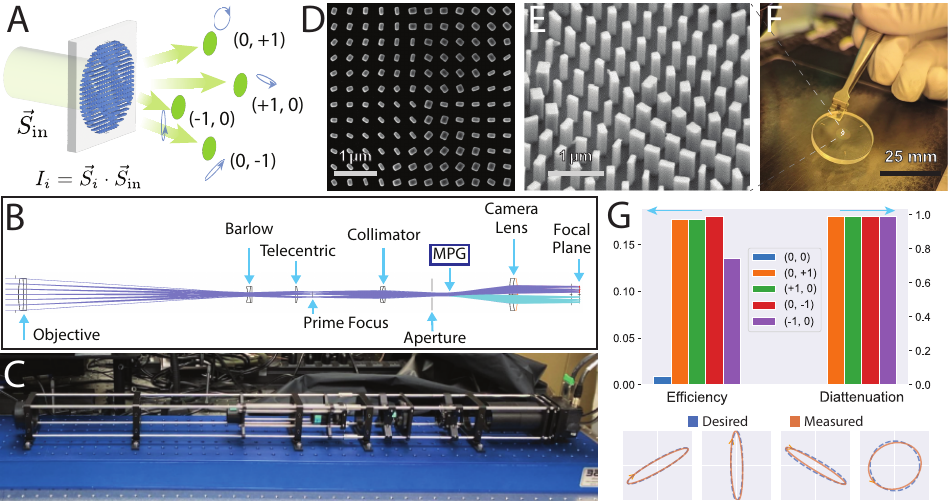}
    \caption{\textbf{MPG and SIMPol design and performance}: \textbf{(A)} Schematic of an MPG, in which incident light whose polarization state is given by Stokes vector $\vec{S}_{\text{in}}$ is distributed primarily amongst four polarization-analyzing orders whose exiting intensities are given by $I_i = \vec{S_i}\cdot\vec{S}_{\text{in}}$ where $\{\vec{S_i}\}$ are the characteristic polarization states of the four orders (illustrated by polarization ellipses). \textbf{(B)} Optical layout and \textbf{(C)} photograph of assembled SIMPol instrument. SIMPol's design provides a pupil plane in which the MPG can be placed, labeled as MPG, before forming four images of a solar scene under investigation. \textbf{(D)} Top-down and \textbf{(E)} oblique scanning-electron microscope (SEM) images of one unit cell of the MPG designed for SIMPol, a photograph of which is shown in \textbf{(F)} as-fabricated on a substrate over a 6 mm diameter. \textbf{(G)} As-characterized performance of the MPG shown in (C-E). The efficiency and diattenuation of each order are shown (top) alongside the polarization states $\{\vec{S}_{i}\}$ for which each order analyzes (bottom), showing close correspondence with the design.}
    \label{fig:SIMPol_and_grating}
\end{figure*}

\subsection{Metasurface Polarization Gratings}

An MPG is a passive, periodically repeating free-space optical element (i.e., a grating) comprised of subwavelength nanostructures exhibiting form birefringence such that the behavior of the grating is polarization-dependent. Using methods of Fourier optics combined with the Jones calculus~\cite{Rubin2019}, an MPG can be designed such that incident polarized light is directed into a number of outgoing diffraction orders according to its polarization state (Fig. \ref{fig:SIMPol_and_grating}(A)). In particular, if the incident light has a polarization state given by the Stokes vector $\vec{S}_{\text{in}}$, the intensity directed to the $i^{\text{th}}$ diffraction order is given by $I_i = \vec{S}_i \cdot \vec{S}_{\text{in}}$, where $\vec{S}_i$ describes each order's analyzed polarization state, efficiency, and polarization contrast (diattenuation). Measurement of $\{I_i\}$ on at least four orders permits reconstruction of $\vec{S}_{\text{in}}$, provided that the $\{\vec{S}_i\}$ of all orders are known (from calibration). In the case of SIMPol, the MPG concentrates incident light into a set of exactly four channels in a way that permits reconstruction of the Stokes vector from a single passive measurement (\emph{snapshot} polarimetry). When the MPG is integrated into an imaging system (ideally in a pupil plane), this polarization analysis occurs in parallel over the system's angular imaging FOV. Four images are formed on the sensor corresponding to these diffraction orders, which can be later analyzed pixel-wise to yield $\vec{S}_{\text{in}}$ over an image. An optimum set of four polarization states can be chosen as the states analyzed by the four inner diffraction orders of an MPG which are maximally distinct from one another and thus optimal for Stokes vector determination~\cite{Rubin2019}. One such example is shown in the four polarization ellipses depicted in Fig. \ref{fig:SIMPol_and_grating}(A).

\subsection{Instrument Design}

The optical design of SIMPol is shown in Fig. \ref{fig:SIMPol_and_grating}(B). The front-end of the system up to its prime focus position is a conventional refractive telescope with a $\sim\diameter 50$ mm objective lens, a Barlow lens to reduce total track length, and a final telecentric re-imaging lens. The back-end of SIMPol is a relay optical system where a second pupil is created and sized appropriately for a $\sim\diameter 6$ mm MPG element (Fig. \ref{fig:SIMPol_and_grating}(B)). A final field lens re-images the diffraction orders produced by the MPG onto a large-format detector permitting snapshot full-Stokes acquisition as described above.

SIMPol's angular FOV is $0.5\degree$ meaning it may function either as a standalone telescope with an FOV sufficient to capture the full solar disk with a nominal spatial resolution of 2.3 arcsec at 460.7 nm or, alternatively, SIMPol can be matched to the exit pupil of a large facility telescope to achieve higher spatial resolution over a reduced FOV, possibly aided by the use of adaptive optics when available.

SIMPol was designed for use at and around 460.7 nm, a wavelength originally chosen due to its proximity to a Sr I absorption line in the solar spectrum which is known for displaying polarized signatures of the Hanle effect~\cite{Zeuner2022}, deemed at the project's start to be a target of interest. This wavelength informs the design of SIMPol's MPG, but the instrument and its design concept could extend to any solar line(s) of interest---SIMPol is constructed from achromatic doublets to allow this flexibility; any change in observational wavelength beyond $\Delta \lambda\sim 20$ nm (see supplement), however, would necessitate use of a different MPG due to the the chromatic dependence of the MPG's diffraction angle and polarimetric performance.

SIMPol is built from custom-designed lenses and has a total length of about 1 meter and is paired with a large-format, 47.5 MPixel focal plane array (FPA) which oversamples SIMPol's diffraction-limited resolution. This 1 meter total track length is comparable to or smaller than other solar polarimeters deployed in observatories and in-orbit (including on CubeSats); the total length was chosen for ease of alignment as a proof-of-concept instrument but could be made shorter if desired (see supplement). A photograph of the SIMPol instrument as-built is shown in Fig. \ref{fig:SIMPol_and_grating}(B), bottom. A full description of SIMPol's design and optical prescription is provided in the supplementary information. There, we also provide several alternative design concepts for SIMPol.

\subsection{Grating Design and Performance}

The design of the MPG used in SIMPol requires special consideration. For one, metasurface elements in general can be difficult for lithographic reasons to scale to large areas. At first glance, this may seem to preclude their use in astronomy which necessitates large apertures for resolution and throughput. This would be especially true if a metasurface is envisioned as playing the role of a telescope's objective lens, as has been implied in past work~\cite{Zhang2023, Park2024, Majumder2025}. However, in SIMPol, the metasurface component's role as an intermediate component of a larger optical system can be used to tradeoff aperture size and angle-of-incidence (as governed by \'etendue conservation): The reimaging (camera) optics on the back-end of SIMPol can be adjusted to control the area of the pupil plane which the MPG must fill at the expense of increased incident angle (AOI) on the MPG and thus its required grating angle. SIMPol's MPG is sized to fill a 6 mm diameter pupil plane with roughly $\pm 5\degree$ AOI. This $\pm5\degree$ angular spread at the MPG then necessitates that the MPG itself deflects incident light by $\pm5\degree$ into its inner diffraction orders to avoid order overlap at the sensor~\cite{Rubin2022}. These specifications (6 mm grating diameter, first orders deflected at $\pm5\degree$) represent a compromise between an area amenable to nanofabrication and a very modest grating deflection angle (equivalently, grating density or period).

These requirements from SIMPol's optical design then inform a design of the required MPG from form-birefringent, subwavelength-sized TiO$_2$ nanostructures, using methods previously reported elsewhere~\cite{Rubin2019, Rubin_Polarization_Grating_Design_2023}. One unit cell of the resultant grating, which consists of a $12 \times 12$ array of structures, occupies a footprint of $4.29\times4.29$ \textmu m. Fig. \ref{fig:SIMPol_and_grating}(D) and (E) show one unit cell of the grating (top down and oblique view SEMs, respectively). Thousands of these unit cells are tessellated along both Cartesian axes to form the final 6 mm MPG for use in SIMPol, as photographed in Fig. \ref{fig:SIMPol_and_grating}(F).

The final grating was tested as a standalone component using two polarimetric testbenches, one in which the MPG is tested alone (for overall efficiency characterization) and one in which it is characterized inside of the SIMPol system. In both cases, the MPG is illuminated using a laser light source of controllable, variable polarization to characterize the polarimetric behavior of each zero and first order. This procedure is described in the supplement. Measurement of the optical power in each order in response to light of variable polarization state permits characterization of three important quantities: (i) Efficiency (defined as the fraction of incident energy directed to an MPG order for incident unpolarized light), (ii) diattenuation (the order's selectivity to polarized light which ranges between 0 and 1, with 0 describing polarization agnostic behavior and 1 describing ideal polarizer-like behavior), and (iii) the polarization state analyzed by each order. All three metrics are contained within $\vec{S}_i$, the analyzer vector for each order which is characterized with these methods.

Fig. \ref{fig:SIMPol_and_grating}(G) summarizes the grating's performance from this measurement. The bar charts report the efficiency and diattenuation of each of the four orders (left and right halves of the plot, respectively). Below, polarization ellipses depict the polarization state for which each order analyzes, showing both the as-fabricated performance (orange, solid) and the predicted design performance (blue, dashed).

The possibility to design and fabricate a working MPG in this way has been shown in our previous work~\cite{Rubin2019, Rubin2022, Li2023}. However, the grating developed here for SIMPol represents a significant improvement in performance with regard to all metrics: This MPG offers a substantially larger aperture (6 mm diameter versus sizes between 1-3 mm in previous work). Its efficiency is markedly higher (67\% total here in the four orders of interest versus 20-45\% in previous works, now far in excess of the theoretical maximum of 50\% offered by a traditional polarizer and on par with or higher than typical reflective blazed gratings in wide use in astronomy and spectroscopy more generally). This is accompanied by significantly lower zero order light leakage ($<$1 \% of incident light directed to the zero order here, while in previous works the zero order was as bright or brighter than the signal diffraction orders) Its orders moreover offer higher diattenuation (no lower than 98\% in each of the four orders and as high as 99.5\% for all but one, versus as low as 85\% in previous work~\cite{Rubin2022}). Finally, the SIMPol MPG shows extremely close correspondence of the realized polarization states versus design (as in Fig. \ref{fig:SIMPol_and_grating}(G)). It is notable that the MPG developed here for SIMPol functions at a bluer wavelength than in these previous works (460 nm here versus 532-870 nm~\cite{Rubin2019, Rubin2022, Li2023}) necessitating accordingly stricter fabrication tolerances to achieve the performance detailed above.

This performance quality was achieved through large area fabrication optimization, an effort we further detail in the supplement. Full details regarding MPG design, fabrication, and testing are additionally provided there.

In addition, we also conducted space qualification testing on this MPG sample, an important consideration for the use of metasurface optical components more broadly in space-borne instrumentation. To the best of our knowledge, this represents the first reported space qualification of a dielectric metasurface to NASA-specified standards. As we fully detail in the supplement, the MPG was polarimetrically tested before and after a regimen that includes thermal cycling and vibration testing to NASA's published standards for low-Earth orbit space qualification by BAE Systems Space \& Mission Systems. (Radiation testing was not performed, however, as the only materials involved here---fused silica and titanium dioxide---were deemed to possess sufficient space heritage and are not known to significantly darken; see supplement). In the supplement, we present results showing that the polarimetric performance of the MPG remained stable, within the precision of our calibration procedure, through each of these testing steps.

\section{Integration with Dunn Solar Telescope}
\label{sec:integration}

\begin{figure}
    \centering
    \includegraphics[width=\columnwidth]{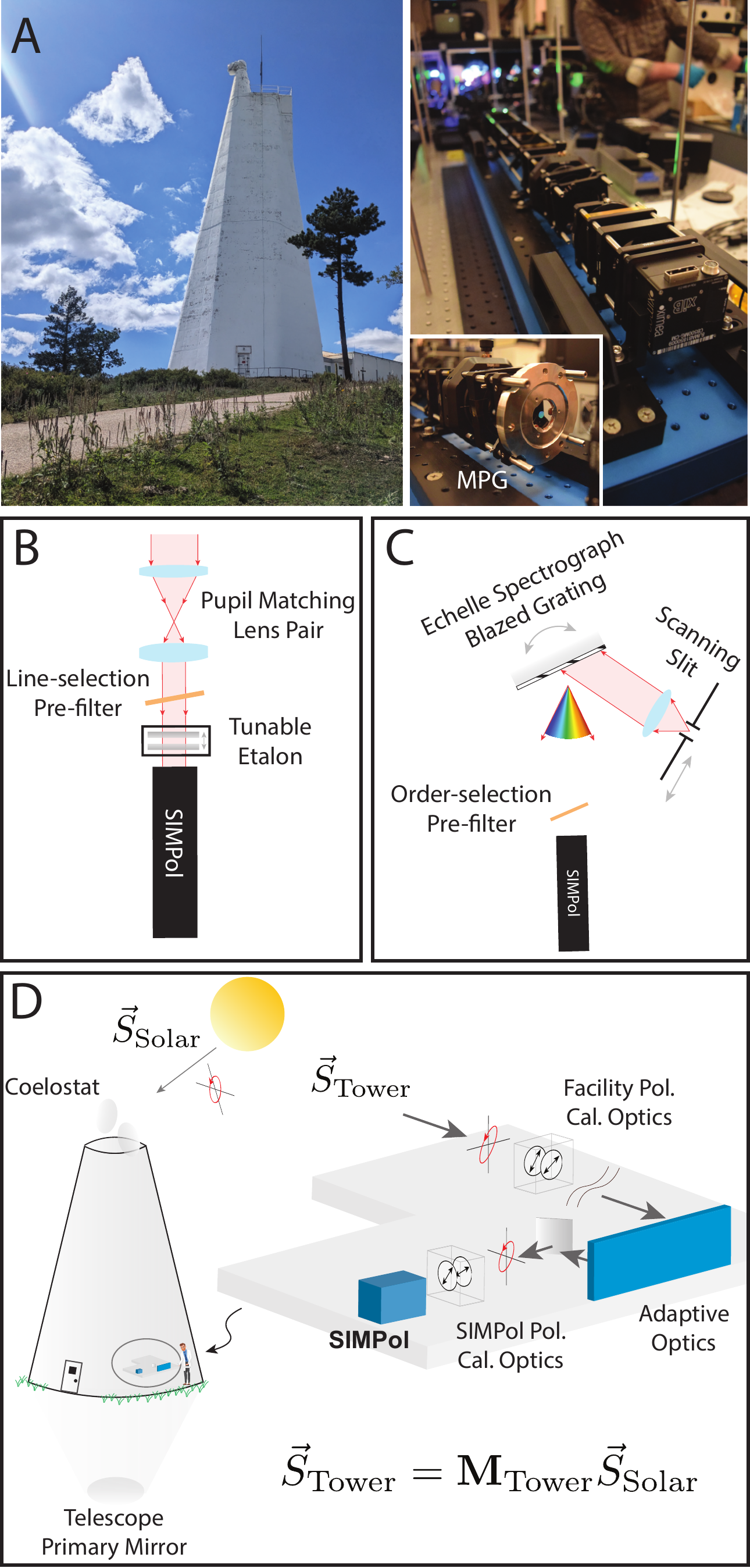}
    \caption{\textbf{Integration with Dunn Solar Telescope (DST)}: \textbf{(A)} View of tower of DST in Sunspot, NM, USA (left). SIMPol operational and pupil-matched to the DST, in the DST's Coud\'e room (right). The MPG integrated within SIMPol (inset). Two SIMPol imaging modes: \textbf{(B)} Schematic of SIMPol used in direct-imaging mode, in which SIMPol is pupil-matched to the DST, and forms 2D Stokes images through a tunable etalon filter. \textbf{(C)} Schematic of SIMPol used in spectrograph mode, in which SIMPol is paired with an on-site echelle spectrograph with a scanning slit. \textbf{(D)} Schematic of setup at DST highlighting important calibration considerations---namely, that the polarization $\vec{S}_{\text{Solar}}^{\prime}$ measured by SIMPol has been modified by the polarization effect of the telescope tower (objective, secondary, and relay mirrors) from the incoming solar light $\vec{S}_{\text{Solar}}$.}
    \label{fig:dst_setup}
\end{figure}

SIMPol, and its enabling MPG element, can acquire full-Stokes polarization imagery of the Sun, even as a standalone instrument. However, SIMPol can additionally be pupil-matched (i.e., its aperture and FOV are matched with lenses) to a larger observatory-grade telescope. This has a number of advantages, most importantly the larger aperture and adaptive optics correction offered by an observatory. Moreover, observatory-integration provides a chance to demonstrate emergent metasurface technology within a much larger scientific-grade facility.

SIMPol was integrated with the Dunn Solar Telescope (DST) located on Sacramento Peak in Sunspot, New Mexico, USA. This facility, the former flagship observatory of the US National Science Foundation's National Solar Observatory, is currently maintained by New Mexico State University for continuing observations and the demonstration of new instrument concepts. The DST consists of a vacuum tower (Fig. \ref{fig:dst_setup}(A), left) atop which is a coelostat (a two-mirror-based solar tracking system) which directs light through a 76.2 cm entrance window aperture to a 1.6 m diameter primary objective mirror located 54 m underground. The primary then relays all sunlight upwards, where it exits the vacuum section through one of six optical windows, after which it is re-collimated by a secondary mirror and directed to the ground-level observing floor. SIMPol is mounted on an optical benchtop on this observing floor, where the DST pupil image can be remapped to SIMPol's entrance pupil via relay lenses and mirrors. Fig. \ref{fig:dst_setup}(A), right, shows SIMPol assembled on an independent breadboard receiving solar light from the exit pupil of the DST. An inset shows the MPG described above mounted within SIMPol.

\subsection{Imaging Modes}

The polarimetric observation of solar magnetism requires spectropolarimetry, that is, wavelength-resolved polarization state measurements of magnetically sensitive spectral lines. SIMPol was used to acquire this data at the DST in two distinct imaging modes. In the first mode, sketched in Fig. \ref{fig:dst_setup}(B), full-Stokes two-dimensional images are acquired at different wavelengths within the usable passband (FWHM nominally was 2.8 pm, but we observed $\sim$9.8 pm in practice) using a tunable Fabry-Perot etalon (FP). We refer to this as the ``Fabry-Perot'' operation mode. Here, an afocal lens pair is used to match SIMPol's entrance pupil to the DST's exit pupil. Spectral sampling is accomplished with a pre-filter (oriented at an angle to select the correct wavelength near the Sr I line of interest) followed by a Fabry-Perot etalon with piezo tuning. In FP mode, four two-dimensional images of the solar scene of interest (corresponding to the MPG's four central diffraction orders) are imaged simultaneously onto SIMPol's detector. The FP etalon can then be scanned over a spectral window of interest to accomplish spectropolarimetry.

We refer to the imaging mode sketched in Fig. \ref{fig:dst_setup}(C) as ``spectrograph'' mode. In this mode, SIMPol serves as a snapshot polarimetry attachment to a traditional grating spectrograph and the image formed on SIMPol's sensor is the polarized spectrum of the region of solar atmosphere spatially sampled by the spectrograph's slit, simultaneously replicated in the four diffraction orders of the polarization-analyzing MPG. Here, SIMPol was combined with the Spectro-Polarimeter for INfrared and Optical Regions (SPINOR)~\cite{Socas2006} which is a permanent facility instrument at the DST. A pupil plane is formed in the proximity of a high-order echelle grating using a collimating lens placed after the slit. The grating has a density of 110 lp/mm, is blazed at $64\degree$, and for our spectral region it was used in order 35. SIMPol receives collimated, spectrally dispersed light from the grating which is passed through a medium ($\sim$10 nm) bandpass pre-filter to prevent diffraction order overlap from the echelle grating. The raw image thus formed by SIMPol contains four spectrally resolved images (with $R\sim10^5$) in four independent polarization states as determined by the MPG, with the $X$ dimension (transverse to the slit) being the spectral dimension, and the $Y$ dimension (along the slit) being the spatial dimension. By scanning the slit over the telescope's FOV along the $X$ dimension, we create a time series of 2D spectro-polarimetric maps for different contiguous slices of the solar region capture by the FOV. In its usual mode of operation, SPINOR would rely on a rotating waveplate retarder in front of the slit to implement a temporal-modulation scheme for polarimetry. This modulator was removed for our experiment, as SIMPol provides metasurface-enabled full-Stokes snapshot polarimetry. We note that the MPG element itself, being a grating, is also spectrally dispersive. However, the spectral dispersion of the spectrograph grating is orders of magnitude larger than that of SIMPol's MPG, so that the spectral dispersion owed to the MPG is deeply sub-pixel in the spectral region of interest of the final image (see supplement).

Significantly more detail about both imaging modes, including detailed optical schematics, is given in the supplemental information.

\subsection{Facility and Calibration} 

In both the FP and spectrograph imaging modes, SIMPol forms four polarization-analyzed images of the solar scene under investigation ($X$-$Y$ images for FP mode, and $\lambda$-$Y$ images for spectrograph mode). In both cases, these images are registered to form (at each pixel across the shared FOV) a four-dimensional vector of intensities $\vec{I}$, from which a Stokes vector can be determined as $\vec{S}=\textbf{O}^{-1}\vec{I}$ where $\textbf{O}$ is a $4\times4$ matrix characterizing the polarization response of each order (with the analyzed order $\{\vec{S}_i\}$ polarization states as its rows). Crucially, the modulation matrix $\textbf{O}$ must be determined pixelwise by a polarization calibration so that its inverse can be used to determine $\vec{S}$. Conducting this polarization calibration in an observatory facility requires special consideration.

In a polarization calibration, one produces an optimized series of input polarization states $\{\vec{S}_\text{in}\}$ while acquiring corresponding output measurements of $\vec{I}$ from the polarimeter to calculate $\textbf{O}$ using a fitting method. This process creates an end-to-end calibration profile of the entire system between where the known input polarization state is generated and the position of the sensor where the output $\vec{I}$ is measured. The sketch of Fig. \ref{fig:dst_setup}(D) shows the complexity of doing so at the DST. Incident polarized light from the Sun arrives at the telescope tower with polarization state $\vec{S}_{\text{Solar}}$, but calibration is conducted with facility polarization optics~\cite{Skumanich1997} which are internal to DST's telescope tower. While these calibration optics allow for polarization calibration of much of the complex optical system which precedes SIMPol on the observatory table (including adaptive optics, various mirrors, and filters), their placement means that the calibration cannot include the combined effects of the coelostat, the primary and secondary mirrors, and other elements within the tower itself. These can induce a polarization effect which is of the same order of magnitude (or larger) than the solar polarized light signal itself. Thus, a Stokes image calculated from any $\mathbf{O}$ calibrated using the facility polarization optics in actuality yields the polarization state $\vec{S}_{\text{Tower}}$ present at the position of the facility calibration optics. This differs from the true solar polarization state $\vec{S}_{\text{Solar}}$ as $\vec{S}_{\text{Tower}} = \textbf{M}_{\text{Tower}}\vec{S}_{\text{Solar}}$. ($\textbf{M}_{\text{Tower}}$ is a $4\times4$ Mueller matrix expressing the polarization effect of the elements within the telescope tower).

Direct measurement of $\textbf{M}_{\text{Tower}}$ would require placement of large aperture polarization optics directly in front of the telescope window atop the tower. At an active, state-of-the-art solar observatory, such a characterization would be conducted regularly; indeed, the Mueller matrix of the DST's tower was characterized regularly in this way in years past~\cite{Skumanich1997, Socas-Navarro2011}. Unfortunately, the most recent telescope Mueller matrix available is not only over 15 years out-of-date, but it additionally does not extend down to SIMPol's operating wavelength.

However, an estimate for $\textbf{M}_{\text{Tower}}$ can be made using a mix of directly measurable data and known physical phenomena. In order to account for the effect of the tower, we can represent the full tower as a series of discrete components~\cite{Lu1996}: $\textbf{M}_{\text{Tower}}=\textbf{M}_{\text{D}}\textbf{M}_{\text{R}}$. $\textbf{M}_{\text{D}}$ is a diattenuator (polarizer-like) Mueller matrix and $\textbf{M}_{\text{R}}$ is retarder (waveplate-like) Mueller matrix. (Here, we assume that the tower exhibits a negligible depolarizing effect). The diattenuating part of the tower $\textbf{M}_{\text{D}}$ can be estimated by observing a quiet (sunspot-free and flare-free) region of the sun where polarization is expected to be small, while the adaptive optics of the facility are set to dither to intentionally blur any residual polarization signatures. These Stokes images are treated as equivalent to observing an unpolarized and featureless source (i.e., $\vec{S}_{\text{Solar}}$ is unpolarized light). The $\vec{S}_\text{Tower}$ observed then corresponds to the characteristic polarization state of the diattenuator $\textbf{M}_{\text{D}}$, which is sufficient to fully reconstruct $\textbf{M}_{\text{D}}$ (full details in supplement). $\textbf{M}_{\text{D}}^{-1}$ can then be applied to SIMPol's measurements to remove this diattenuating effect.

Unfortunately, there is no such method to directly measure the retarding component $\textbf{M}_{R}$. Uncorrected retardance effects from the tower induce a lossless mixing of the different components of the Stokes vector, i.e., a ``polarization crosstalk''. Nonetheless, $\textbf{M}_{R}$ is defined by just three free parameters for a fixed configuration of the telescope. As we show below, given the known polarization signature of magnetic effects being observed (specifically, the Zeeman effect), an estimate for these three parameters (and thus, for $\textbf{M}_{R}$) can be made. In the results that follow, we then apply this tower retardance correction $\textbf{M}_{R}^{-1}$ to the results from both FP and spectrograph mode data. While this estimation step of $\textbf{M}_{R}$ presents an inconvenience and challenge in data reduction, we note that this is a universal issue of polarimetry which is not a consequence of the instrument itself nor the metasurface technology being demonstrated here---one which could be appropriately addressed in a science mission with a dedicated use-case. Exact numerical details concerning tower effect corrections and the specific parameters used are provided in the supplement.

Finally, we note that (as shown in Fig. \ref{fig:dst_setup}(D)) polarization calibration optics can also be placed in front of SIMPol itself, which can help to characterize the polarization effect of the optics preceding SIMPol (see supplement).

\begin{figure*}[htbp]
    \centering
    \includegraphics[width=\textwidth]{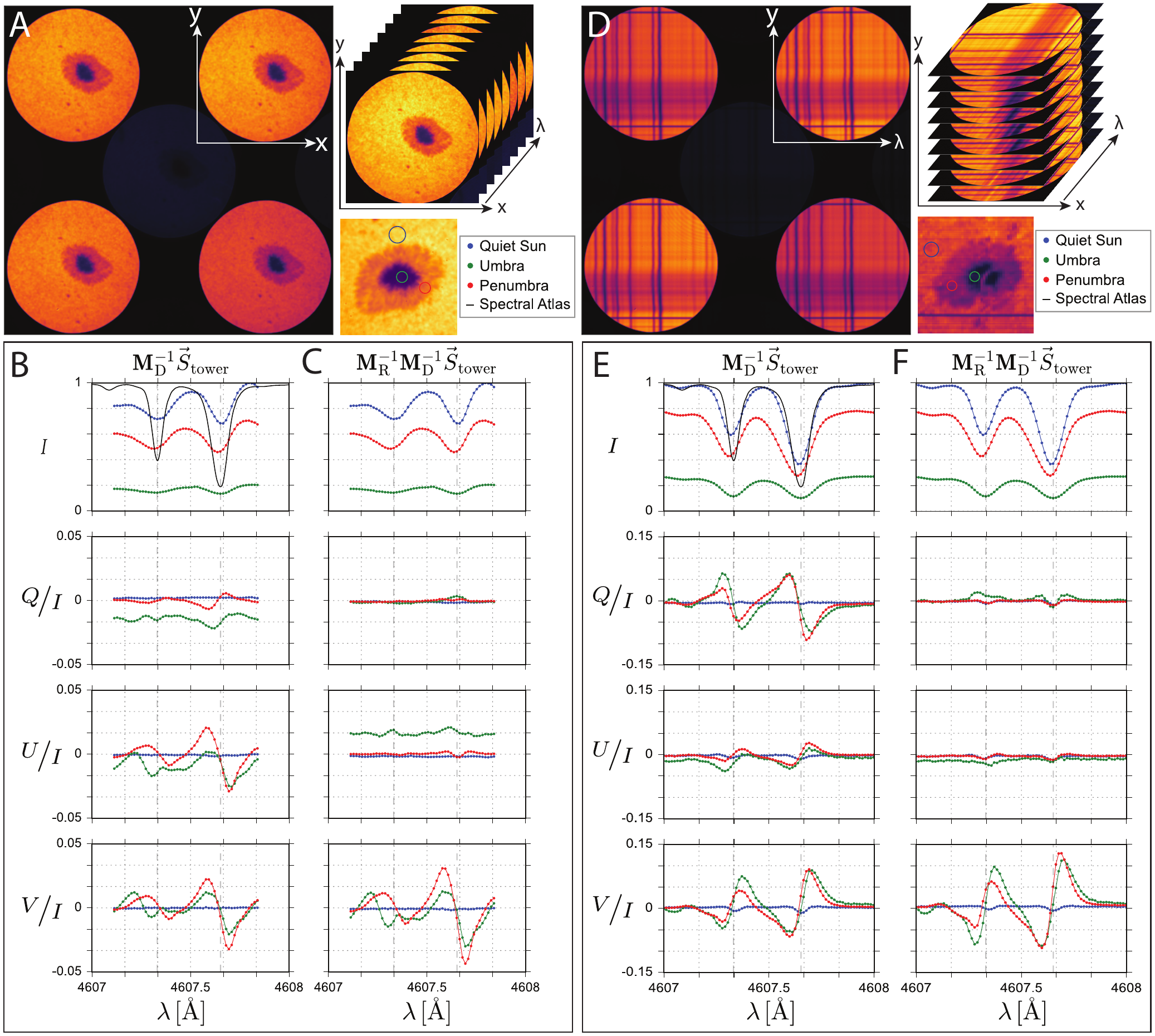}
    \caption{\textbf{Observations of Solar Magnetism with SIMPol}:  \textbf{(A)} Example raw image of sunspots acquired by SIMPol in Fabry-Perot mode with four images of the sunspot feature corresponding to the four inner polarization-analyzing orders of the MPG. A significantly darker zero order is barely visible. Each image represents a 2D X/Y image. A scanning Fabry-Perot permits acquisition of a X/Y/$\lambda$ full-Stokes polarimetric dataset (top right). An image of the sunspot (bottom right) provides a key for locations referenced in (B) and (C). \textbf{(B)} $\vec{S}(\lambda)$ measured in sunspots using SIMPol's Fabry-Perot imaging mode. Each row denotes one component of the Stokes vector ($I$ in arbitrary units normalized to the maximum value displayed, with normalized Stokes components $Q/I$, $U/I$, and $V/I$ in the subsequent three rows). (B) reports the Stokes vector with the measured diattenuation correction $\textbf{M}_{\text{D}}^{-1}$ applied. \textbf{(C)} presents the cross-talk corrected version of the data presented in (b) by including a `best-estimate' retardance correction $\textbf{M}_{\text{R}}^{-1}$. Each curve is taken as the average over a location in the image denoted in (A). $I$ shows the adjacent Sr I and Fe I with spacing and position compared to a solar atlas. Evidence of the longitudinal and transverse Zeeman effect is present. \textbf{(D)} Example raw image of a sunspot from SIMPol's spectrograph imaging mode. Here, each MPG order contains a $Y/\lambda$ spatial/spectral image. Scanning the spectrograph slit permits acquisition of a full $X/Y/\lambda$ dataset (top right). A map of the sunspot formed by scanning slit denotes locations referenced in (E) and (F). \textbf{(E)} $\vec{S}(\lambda)$ measured in SIMPol's spectrograph mode. \textbf{(F)} is the same data passed through an additional processing step using our best-estimate retardance correction. The layout and interpretation of (E) and (F) is identical to that in (B) and (C). Clear (and comparably cleaner) evidence of the longitudinal and transverse Zeeman effect is evident in the $V$ and linear ($Q$ and $U$) components, respectively. Despite their visual similarity, the observed sunspots in each operational mode ((B) and (C) versus (E) and (F)) are from different days and have oppositely oriented magnetic fields as demonstrated in the change in orientation of $V/I$ signals.}
    \label{fig:SIMPol_Results}
\end{figure*}

\subsection{Observations of Solar Magnetism}

Enabled by the MPG, SIMPol can acquire snapshot polarimetric observations of solar magnetism in sunspots. The principal results of these observations are shown in Fig. \ref{fig:SIMPol_Results} with FP and spectrograph operational mode results occupying the left and right halves of the figure, respectively. The FP and spectrograph data shown in Fig. \ref{fig:SIMPol_Results} were acquired on September 13, 2024 (16:01:43 - 16:08:49 UTC) and September 18, 2024 (16:07:18 - 16:23:50 UTC), respectively. Each was processed using calibration datasets acquired the same day. Despite their visual similarity, the sunspots imaged in the two different modes were not the same.

To provide context, we first remark on precisely which observational polarimetric signatures of solar magnetism are expected. The wavelength-resolved Stokes vector $\vec{S} = \begin{bmatrix}
I(\lambda), Q(\lambda), U(\lambda), V(\lambda)\end{bmatrix}^T$ is measured. The primary physical effects on the sun giving rise to polarization in sunlight are the Zeeman and Hanle effects, both of which have similar quantum mechanical origins given the modification of the electron wavefunction in the presence of a magnetic field. The Sr I absorption line at 460.7 nm is known to display this effect~\cite{Zeuner2022}, a fact which originally motivated this as the choice of SIMPol's target wavelength. However, polarization due to the Hanle effect manifests at a sub-percent level in the presence of relatively weak fields ($\sim10^1$-$10^2$ gauss) and can be challenging to observe even for mature, established instrumentation which do not suffer from the observatory calibration ambiguity described above. On the other hand, the Zeeman effect can attain a substantially larger magnitude (single to tens of percent) when observed in strong solar active regions such as sunspots.

There are two manifestations of this Zeeman effect, one which is owed to the projection of the magnetic field along the line-of-sight of observation (the longitudinal Zeeman effect) and one to the projection of the magnetic vector onto the plane perpendicular to the line-of-sight (the transverse Zeeman effect). In both cases, and indeed even in the absence of magnetism or polarization $I(\lambda)$ exhibits dips characteristic of solar absorption features which are roughly Gaussian in shape (more precisely, a Voigt profile). If a longitudinal magnetic field is present, $V(\lambda)$ will tend to take on an anti-symmetric shape with respect to line center, with amplitude proportional to the derivative of the intensity, $\partial I/\partial \lambda$ if the field strength is not so large to produce a clear Zeeman broadening or splitting of the intensity profile (the so-called ``weak field approximation''). The scaling factor depends linearly on the strength of the line-of-sight magnetic field, and on the atomic properties of the spectral line. If a transverse magnetic field is present, a symmetric variation in $Q(\lambda)$ and $U(\lambda)$ is produced instead, whose shape is approximately given by the second derivative of the intensity profile,  $\partial^2 I/\partial\lambda^2$. Its amplitude depends quadratically on the transverse $B$ field, making this a second-order effect that is generally weaker and consequently more difficult to observe.

\subsection{Fabry-Perot Mode}

In Fabry-Perot mode, SIMPol produces a direct image of the Sun at the sensor. Fig. \ref{fig:SIMPol_Results}(A) shows a raw image acquired by SIMPol in FP mode for a fixed setting of the FP etalon. Four bright central diffraction orders are present in the image, each capturing an image of a sunspot effectively analyzed with respect to a different polarization state. This sample raw image corroborates the MPG performance outlined in Fig. \ref{fig:SIMPol_and_grating}(G): the four first orders are approximately equally bright and contain significantly more energy than the barely visible central zero order. (This raw image shows that this high performance is maintained even when the entire 6 mm grating aperture is illuminated in the pupil plane of the instrument). Given a raw image (e.g., Fig. \ref{fig:SIMPol_and_grating}(A)), the inner orders can be registered and a calibration $\textbf{O}$ applied pixel-wise to obtain a polarization image of $\vec{S}(\lambda)$. The etalon can be scanned to assemble a wavelength-resolved Stokes vector image over variable $\lambda$, as implied by the graphic in the top right of Fig. \ref{fig:SIMPol_Results}(A). Regions of the resulting Stokes image can be selected and averaged in order to view a one-dimensional plot of $\vec{S}(\lambda)$---a few chosen regions are denoted in the bottom left of Fig. \ref{fig:SIMPol_and_grating}(A), namely a quiet region of the sun outside the sunspot, as well as the sunspot's penumbra and central umbra.

Figs. \ref{fig:SIMPol_Results}(B) and (C) present each element of the wavelength-resolved Stokes vector from each region and show the results of the Stokes vector at two stages of the tower correction described above. Fig. \ref{fig:SIMPol_Results}(B) reports results with only tower diattenuation correction applied ($\textbf{M}_D^{-1}$), while Fig. \ref{fig:SIMPol_Results}(C) includes an additional best-estimate retardance correction ($\textbf{M}_R^{-1}\textbf{M}_D^{-1}$, yielding an estimate of the true $\vec{S}_{\text{Solar}}(\lambda)$) with tower effects removed. A black curve ($I$ in left column only) gives the expected intensity referenced from a solar atlas, showing two absorption lines between 460.7-460.8 nm attributed to Sr I (left) and Fe I (right) features whose center wavelengths ($\lambda_\text{Sr I}$ = 460.73 nm and $\lambda_\text{Fe I}$ = 460.76 nm) are marked in each plot by black dashed lines.

The plots of $Q/I$, $U/I$, and $V/I$ in Figs. \ref{fig:SIMPol_Results}(B) and (C) show an anti-symmetric profile characteristic of the longitudinal Zeeman effect. This effect vanishes in the quiet (non-sunspot) region of the image denoted by the blue curve, as expected from both the weakening of the magnetic strength and the increasingly inclined field outside the sunspot, which, for the geometry of this observation, makes the field mostly transversal to the LOS. However, this anti-symmetric line profile is expected to only appear in $V(\lambda)$. Its presence in the linear Stokes components suggests significant crosstalk which can be ascribed to retardance from the elements within the telescope tower. In Fig. \ref{fig:SIMPol_Results}(C), on the other hand, the retardance correction $\mathbf{M}_\text{R}^{-1}$ correctly transfers most of this spurious signature in the linear components ($Q/I$ and $U/I$) to the circular Stokes component ($V/I$). In the process, evidence of the transverse Zeeman effect signature, with its characteristic symmetric Zeeman triplet shape (which was previously obscured by the much larger cross-talk signal). Particularly in the penumbra (red curve) for the Fe I line, variations in $Q/I$ and $U/I$ are observed that are symmetric about line center and shaped as the second derivative of the line profile ($\partial^2I/\partial\lambda^2$).

We note that in these results it is evident that the spectral lines are blurred by limited spectral resolution, a consequence of the FWHM of the specific FP etalon on-hand at the DST (see supplement) and is not reflective of the MPG technology being demonstrated; no particular attempt has been made to spectrally de-convolve the FP profile here. Moreover, we note that for each region, the lines and Zeeman signatures are centered at slightly different locations; this stems from the expected angular dispersion effects induced by the etalon. We also believe that this data (particularly the $U/I$ trace from the umbra, in green) shows there are residual crosstalk and polarization aberration effects which could not be accounted for in the current method of estimating the effects using the $\mathbf{M}_\text{Tower}\approx\mathbf{M}_\text{R}\mathbf{M}_\text{D}$ approximation, but would be characterized and correctable in a direct measurement of the $\mathbf{M}_\text{Tower}$. Finally, the etalon itself presents a delicate challenge of optical alignment.

Using SIMPol in an FP mode of imaging presents several advantages - not least of which is the ability to simultaneously acquire the full 2D context of solar features as in Fig. \ref{fig:SIMPol_Results}(A). However, the use of a scanning etalon for imaging presents a number of challenges (some of which we remarked on above). More fundamentally, with an etalon both space dimensions are acquired simultaneously at the expense of the spectral profile which must be built up in time. It is crucially this spectral profile (of the Stokes vector) which is necessary for the remote sensing of the magnetic field. This is a classic tradeoff in astronomical and solar instrumentation generally~\cite{Iglesias2019}, one which can be solved by use of a spectrograph instead which permits simultaneous spectral acquisition at the expense of one space dimension. In the next section, we show that SIMPol and its MPG can equally well be used in such a spectrometer architecture.

The FP mode acquires 2D images at the expense of simultaneous acquisition of spectra which must be measured over time by scanning an etalon. This limits time resolution in spectro-polarimetric applications (such as the polarimetric determination of solar magnetic fields). This can be solved by use of a spectrograph instead which permits simultaneous spectral acquisition at the expense of one space dimension. In the next section, we show that the snapshot polarimetric advantage offered by SIMPol and its MPG can be also leveraged within a spectrograph.

\subsection{Spectrograph Mode}

In a spectrograph, the spectral content of a signal can be acquired simultaneously. Paired with the MPG in SIMPol, this permits snapshot determination of the wavelength-resolved Stokes vector needed to reduce solar magnetic fields. While this sacrifices the ability to acquire 2D spatial data simultaneously, this information is not always needed and can be recovered by scanning the spectrograph's slit. Simultaneous spectroscopic measurements such as these---both polarization-sensitive and agnostic---are often preferred in solar astronomy, in particular for diagnostic measurements of solar plasma under fast variation (e.g., in the presence of strong and rapidly changing velocity fields).

Fig. \ref{fig:SIMPol_Results}(D) shows a raw frame from SIMPol integrated with a grating spectrometer as sketched in Fig. \ref{fig:dst_setup}(C) above. The X-Y-$\lambda$ raw frames in Fig.~\ref{fig:SIMPol_Results}(D) depict how the same information is now collected in an alternate format. Again, the innermost four orders are observed to be approximately equally bright with a significantly dimmer zero order. Each order gives a spatially-resolved spectrum such that pixel columns span spatial sunspot information while pixel rows contains spectral data spanning 460.4-461.0 nm. The Sr~I and Fe~I lines of interest are the central-most line pair in each MPG order. Along the vertical direction, the image slices through a sunspot. The lines can be observed to broaden and shift in response to the Doppler effect from moving plasmas in these locations. From the 2D image of the observed sunspot that is formed after a slit scan, different regions in the quiet sun, umbra, and penumbra can be selected for analysis.

Figs. \ref{fig:SIMPol_Results}(E) and (F) give the wavelength-resolved Stokes vector in each selected region, Fig.\ref{fig:SIMPol_Results} (E) reports $\mathbf{M}^{-1}_\text{D}$ diattenuation-only processed spectral data and Fig\ref{fig:SIMPol_Results}(F) reports the same data processed using the best-estimate $\mathbf{M}^{-1}_\text{R}\mathbf{M}^{-1}_\text{D}$ tower retardance and diattenuation corrections applied. In this use case, the sunspot data captured in spectrograph mode  yields much finer spectral sampling and higher spectral resolution than the resolution seen in the prior section's implementation of FP-mode. This is expected considering the aforementioned limitations of our etalon filter. Again, strong evidence of the Zeeman effect is present as an anti-symmetric, first-derivative-shaped profile about line center in the umbral and penumbral regions, while remaining below detection in the neighboring quiet-sun atmosphere. The presence of the same signature in $Q/I$ and $U/I$ is again due to crosstalk induced by the tower's retardance. A best-estimate attempt at removing this effect in the right half of (D) again transfers this signature of the longitudinal Zeeman effect into $V/I$ and again reveals---significantly more clearly this time---the physically expected symmetric, second derivative-like line profiles in $Q/I$ and $U/I$ that are characteristic of the transverse Zeeman effect for transverse B fields. These are very clearly evident for both spectral lines in the umbra and penumbra, whereas they are obscured by crosstalk in the left column of (D).

These Stokes profiles are readily recognizable as evidence of the Zeeman effect, here acquired in a completely novel and advantageous way by a nanophotonic polarization technology integrated with established astronomical instrumentation.

\subsection{Metasurface Magnetography}

The Zeeman-induced polarized line shapes of Fig. \ref{fig:SIMPol_Results}(F) from SIMPol used in spectrograph mode can be used to provide a quantitative estimation of the magnetic field strength in the observed sunspot region. In the so-called ``weak-field approximation'', measured profiles for $V/I$ can be fit to a finite-differenced form of $\partial I/\partial \lambda$; the coefficient of best fit, combined with the effective Land\'e $\overline{g}$ of a given line, gives the spatially-resolved, line-of-sight magnetic field\footnote{$\overline{g}$ is a linear combination of the usual Land\'e factors of the upper and lower levels of the atomic transition corresponding to the observed spectral line~\cite{Degl2006}.}. This is a standard technique of solar polarimetry~\cite{Degl2006}, and is described further in the supplement. If this is done for each point in the spectral image formed by scanning the spectrograph slit, SIMPol can be used as a line-of-sight (longitudinal) solar magnetograph---reporting a solar observable in physical units. 

\begin{figure}[!t]
    \centering
    \includegraphics[width=\columnwidth]{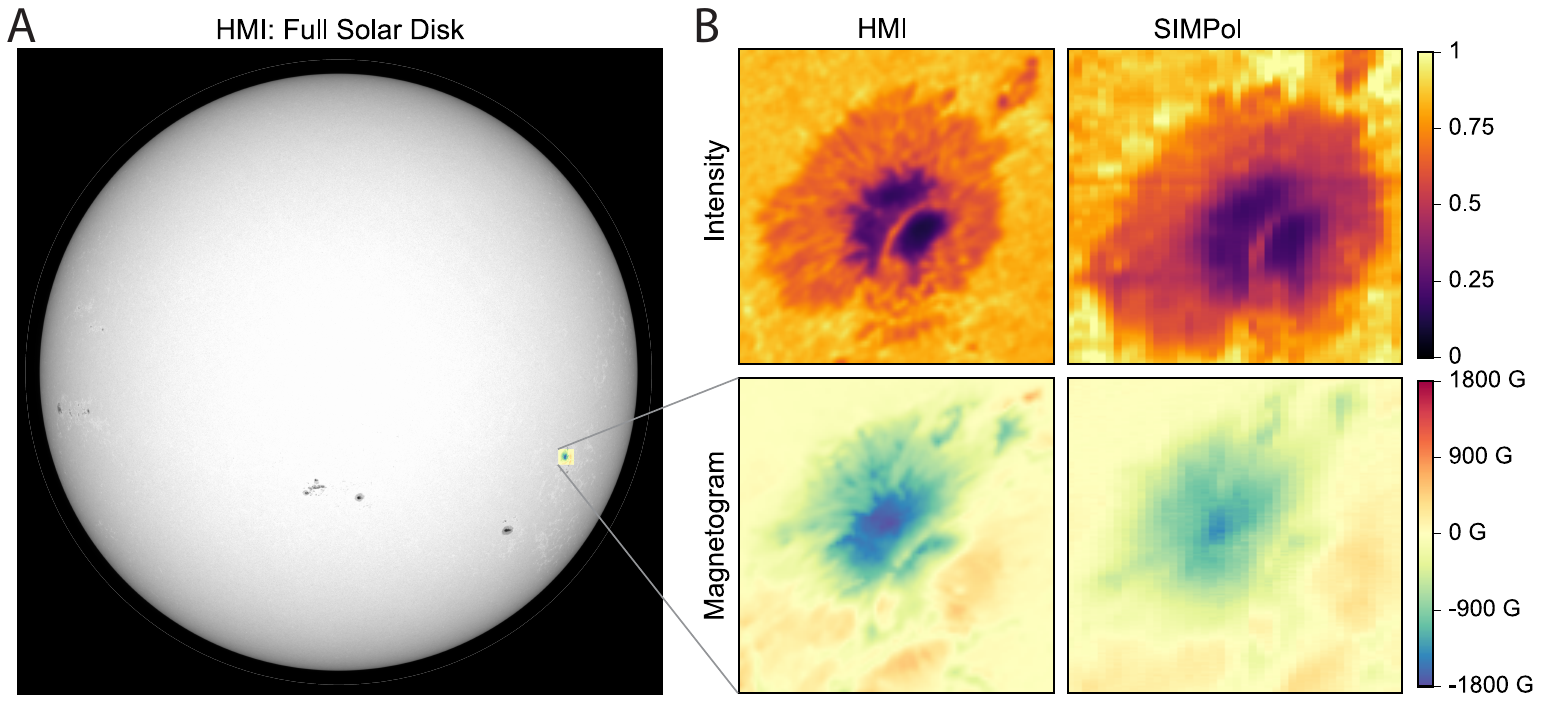}
    \caption{\textbf{Longitudinal Magnetic Fields}: Comparison of the intensity (top) and longitudinal magnetic field in Gauss (bottom) derived from Stokes $V(\lambda)$. The result obtained from SIMPol (right column) is compared to that from the Helioseismic and Magnetic Imager (HMI) for observations made at 14:00 UTC on September 18th, 2024. The location of the observed sunspot is indicated as an inset on the full solar disk as imaged by HMI (left). The Fe I line (460.76 nm) was used to calculate the SIMPol results presented here. A rotation of 35$\degree$ was applied to the HMI data to better match the SIMPol Grating Spectrometer mode image. Negative values correspond to fields pointing toward the sun.}
    \label{fig:HMI_vs_SIMPol}
\end{figure}

This is shown in the rightmost column of Fig. \ref{fig:HMI_vs_SIMPol}, where the top shows the normalized intensity $I$ observed by SIMPol over the sunspot, while the bottom shows the line-of-sight magnetic field (in units of gauss) deduced from the measured $V/I$. (Negative values correspond to fields pointing toward the sun). The Fe I line as observed by SIMPol is used in deducing the strength of the LOS magnetic field, because if its somewhat higher ($\sim$25\%) magnetic sensitivity than the Sr~I line.

This magnetogram can then be compared to data acquired by the Helioseismic and Magnetic Imager (HMI), a polarimetric instrument aboard NASA's Solar Dynamics Observatory. Fig. \ref{fig:HMI_vs_SIMPol}(A) shows a view of the Sun acquired by HMI at the same time (within 14 minutes) as the observations acquired with SIMPol at the DST in spectrograph mode. The same sunspot is identified, and the intensity and line-of-sight magnetic field in gauss as determined by HMI are shown in the left column of Fig. \ref{fig:HMI_vs_SIMPol}(B).

Direct comparison of HMI and SIMPol's spectrograph results is complicated by the fact that HMI and SIMPol are quite different as instruments: HMI acquires its polarization information with a combination of a rotating waveplate and a tunable filter---by acquiring full 2D images rather than relying on a moving slit as we have, HMI's data offers significantly higher resolution here. Nonetheless, the two compare favorably. SIMPol produces the correct order of magnitude line-of-sight magnetic fields, and many of the fine features of the sunspot's magnetic landscape from HMI are corroborated by SIMPol. For example, the hook-like feature on the top right of the sunspot, and the regions of plage beneath the sunspot where the field points oppositely to the field about the umbra (colored red in both magnetograms).

The two magnetograms in Fig. \ref{fig:HMI_vs_SIMPol}(B) do display a systematic discrepancy in determined field (the maximum magnitude reported by HMI is roughly -1750 G, versus -1450 G for SIMPol). In the supplement, we discuss the origins of this discrepancy which, we believe, stem from calibration and data reduction challenges inherent to this preliminary demonstration.

%A discussion on the details behind the detailed discrepancies between the HMI and SIMPol magnetograms is provided in the supplement.

%A slight discrepancy exists between the line-of-sight fields reported by HMI and those inferred with SIMPol. The maximum magnitude reported by HMI is roughly -1750 G, versus -1450 G for SIMPol. We attribute this to three factors: 1) SIMPol's limited resolution versus HMI, which may decrease fields near the observed peak in the umbra by blurring. 2) Uncertainty on the exact Land\'e $\overline{g}$ factor to be used for the Fe I line, which is reported to be blended with another line. 3) A depression of the $V/I$ signal observed with SIMPol because of subtle residual image aberrations which differ between the MPG orders during data reduction. Applying only the simplest image registration strategy based on linear shifting of the sub-images by an integer number of pixels is likely to leave some misalignment uncorrected; this has the effect of lowering the magnitude of the observed polarization and calculated magnetic field. We describe in the supplement how registration involving nonlinear transformations of each sub-image can improve this and produce larger magnitudes of observed polarization. This issue, and the nonlinear image registration techniques used to fix it, are well known in solar instrumentation, and are not a challenge unique to SIMPol or the MPG technology.

\section{Conclusion}
\label{sec:conclusion}

In recent years, metasurface optical elements have seen intensive academic and industrial interest, primarily for their potential in miniaturizing free-space optical systems of all kinds with flat, multifunctional components. The ability to spatially structure light's polarization state is perhaps the foremost advantage of metasurface optics above more conventional optics and other diffractive components of the past. All of these advantages are particularly relevant for astronomy, both ground-based and space-borne, a field which has historically driven advances in optical technology. In this work, we have demonstrated an optical system in which a metasurface serves as an enabling element in a new type of astronomical instrument, in this case a polarimeter for the snapshot characterization of solar magnetism. We designed an instrument around a metasurface polarization grating, developed a high-performing version of this MPG which was environmentally qualified for space launch, and subsequently integrated this MPG into SIMPol which was then deployed to a major solar observatory. We have additionally presented preliminary scientific results showing its ability to observe the Zeeman effect in sunspots and produce a quantitative mapping of longitudinal magnetic fields in sunspots using these observations. To the best of our knowledge, this represents one of the first demonstrated uses of a metasurface in an on-sky astronomical observatory science instrument.

In early academic, component-level research, metasurfaces were often envisioned as replacing entire optical systems with a single component. In this work, in contrast, a metasurface is just one part of a complex multi-element system consisting mostly of traditional mirrors, lenses, and filters (indeed, including the entire Dunn Solar Telescope beampath). The metasurface plays one specific role which leverages its ability to control polarized light, the functionality which most distinguishes metasurface components from other classes of optics. In this way, the present work represents a maturation of metasurfaces and their conceived role in astronomical instrumentation and optical systems more generally.

Recent years have also seen significant interest in ``astrophotonics''~\cite{Jovanovic2023}---advanced optical technologies, especially nanotechnologies, for astronomy. Many of these technologies (e.g., arrayed waveguide grating spectrometers~\cite{Cvetojevic2012, Gatkine2017} or photonic lanterns~\cite{Bland2012, Leon2013, Moraitis2021}), are guided-wave devices either on a single-mode integrated photonic chip or in optical fiber. This makes their deployment in astronomical systems (which are inherently free-space and highly multimode) extremely challenging. In contrast, the MPGs demonstrated here are a new (nano)technology for astronomical observations which operate in free-space, thus easing their integration into and cooperation with existing astronomical systems more broadly.

Further work will refine SIMPol and its calibration and data processing pipeline (see supplement). We envision that instruments like SIMPol, enabled by metasurface components, may play an important role in future solar observatories. We see particular promise for the ability to merge snapshot metasurface polarimetry with slit spectrograph magnetometry of solar lines. Current spectrographic solar polarimeters (including the Visible Imaging Spectropolarimeter~\cite{DeWijn2022} currently operational at the Daniel K. Inouye Solar Observatory in Hawai`i) could be enhanced by the inclusion of metasurface components in the future. We believe the type of single-element, snapshot solar polarimetry demonstrated here will moreover benefit future spaceborne polarimetric missions, where passive polarimetry free of moving components is highly desirable. Brief commentary on future instrument architectures which may be enabled by the work presented here is provided in the supplement.

\section{acknowledgments}

This work was supported by the Heliophysics Technology and Instrument Development for Science (HTIDeS) program of the Heliophysics Strategic Technology Office (HESTO) of the National Aeronautics and Space Administration (NASA) under grant number 80NSSC21K0799. This work was additionally performed in part at the Harvard University Center for Nanoscale Systems (CNS), a member of the National Nanotechnology Coordinated Infrastructure Network (NNCI), which is supported by the United States National Science Foundation (NSF). Solar observations in this work were conducted at the Dunn Solar Telescope, a part of the Sunspot Solar Observatory, which is maintained by New Mexico State University on behalf of the NSF. We thank Kevin Reardon (National Solar Observatory) for the loan of the tunable etalon used in this work.

%\bibliography{Noah-Lisa}% Produces the bibliography via BibTeX.

\begin{thebibliography}{36}%
\makeatletter
\providecommand \@ifxundefined [1]{%
 \@ifx{#1\undefined}
}%
\providecommand \@ifnum [1]{%
 \ifnum #1\expandafter \@firstoftwo
 \else \expandafter \@secondoftwo
 \fi
}%
\providecommand \@ifx [1]{%
 \ifx #1\expandafter \@firstoftwo
 \else \expandafter \@secondoftwo
 \fi
}%
\providecommand \natexlab [1]{#1}%
\providecommand \enquote  [1]{``#1''}%
\providecommand \bibnamefont  [1]{#1}%
\providecommand \bibfnamefont [1]{#1}%
\providecommand \citenamefont [1]{#1}%
\providecommand \href@noop [0]{\@secondoftwo}%
\providecommand \href [0]{\begingroup \@sanitize@url \@href}%
\providecommand \@href[1]{\@@startlink{#1}\@@href}%
\providecommand \@@href[1]{\endgroup#1\@@endlink}%
\providecommand \@sanitize@url [0]{\catcode `\\12\catcode `\$12\catcode `\&12\catcode `\#12\catcode `\^12\catcode `\_12\catcode `\%12\relax}%
\providecommand \@@startlink[1]{}%
\providecommand \@@endlink[0]{}%
\providecommand \url  [0]{\begingroup\@sanitize@url \@url }%
\providecommand \@url [1]{\endgroup\@href {#1}{\urlprefix }}%
\providecommand \urlprefix  [0]{URL }%
\providecommand \Eprint [0]{\href }%
\providecommand \doibase [0]{https://doi.org/}%
\providecommand \selectlanguage [0]{\@gobble}%
\providecommand \bibinfo  [0]{\@secondoftwo}%
\providecommand \bibfield  [0]{\@secondoftwo}%
\providecommand \translation [1]{[#1]}%
\providecommand \BibitemOpen [0]{}%
\providecommand \bibitemStop [0]{}%
\providecommand \bibitemNoStop [0]{.\EOS\space}%
\providecommand \EOS [0]{\spacefactor3000\relax}%
\providecommand \BibitemShut  [1]{\csname bibitem#1\endcsname}%
\let\auto@bib@innerbib\@empty
%</preamble>
\bibitem [{\citenamefont {Dorrah}\ and\ \citenamefont {Capasso}(2022)}]{Dorrah2022}%
  \BibitemOpen
  \bibfield  {author} {\bibinfo {author} {\bibfnamefont {A.~H.}\ \bibnamefont {Dorrah}}\ and\ \bibinfo {author} {\bibfnamefont {F.}~\bibnamefont {Capasso}},\ }\bibfield  {title} {\bibinfo {title} {Tunable structured light with flat optics},\ }\bibfield  {journal} {\bibinfo  {journal} {Science}\ }\textbf {\bibinfo {volume} {376}},\ \href {https://doi.org/10.1126/science.abi6860} {10.1126/science.abi6860} (\bibinfo {year} {2022})\BibitemShut {NoStop}%
\bibitem [{\citenamefont {Rubin}\ \emph {et~al.}(2021)\citenamefont {Rubin}, \citenamefont {Shi},\ and\ \citenamefont {Capasso}}]{Rubin2021a}%
  \BibitemOpen
  \bibfield  {author} {\bibinfo {author} {\bibfnamefont {N.~A.}\ \bibnamefont {Rubin}}, \bibinfo {author} {\bibfnamefont {Z.}~\bibnamefont {Shi}},\ and\ \bibinfo {author} {\bibfnamefont {F.}~\bibnamefont {Capasso}},\ }\bibfield  {title} {\bibinfo {title} {Polarization in diffractive optics and metasurfaces},\ }\href {https://doi.org/10.1364/AOP.439986} {\bibfield  {journal} {\bibinfo  {journal} {Adv. Opt. Photon.}\ }\textbf {\bibinfo {volume} {13}},\ \bibinfo {pages} {836} (\bibinfo {year} {2021})}\BibitemShut {NoStop}%
\bibitem [{\citenamefont {Mueller}\ \emph {et~al.}(2017)\citenamefont {Mueller}, \citenamefont {Rubin}, \citenamefont {Devlin}, \citenamefont {Groever},\ and\ \citenamefont {Capasso}}]{Mueller2017}%
  \BibitemOpen
  \bibfield  {author} {\bibinfo {author} {\bibfnamefont {J.~P.~B.}\ \bibnamefont {Mueller}}, \bibinfo {author} {\bibfnamefont {N.~A.}\ \bibnamefont {Rubin}}, \bibinfo {author} {\bibfnamefont {R.~C.}\ \bibnamefont {Devlin}}, \bibinfo {author} {\bibfnamefont {B.}~\bibnamefont {Groever}},\ and\ \bibinfo {author} {\bibfnamefont {F.}~\bibnamefont {Capasso}},\ }\bibfield  {title} {\bibinfo {title} {Metasurface polarization optics: Independent phase control of arbitrary orthogonal states of polarization},\ }\href {https://doi.org/10.1103/PhysRevLett.118.113901} {\bibfield  {journal} {\bibinfo  {journal} {Physical Review Letters}\ }\textbf {\bibinfo {volume} {118}},\ \bibinfo {pages} {113901} (\bibinfo {year} {2017})}\BibitemShut {NoStop}%
\bibitem [{\citenamefont {Arbabi}\ \emph {et~al.}(2015)\citenamefont {Arbabi}, \citenamefont {Horie}, \citenamefont {Bagheri},\ and\ \citenamefont {Faraon}}]{Arbabi2015}%
  \BibitemOpen
  \bibfield  {author} {\bibinfo {author} {\bibfnamefont {A.}~\bibnamefont {Arbabi}}, \bibinfo {author} {\bibfnamefont {Y.}~\bibnamefont {Horie}}, \bibinfo {author} {\bibfnamefont {M.}~\bibnamefont {Bagheri}},\ and\ \bibinfo {author} {\bibfnamefont {A.}~\bibnamefont {Faraon}},\ }\bibfield  {title} {\bibinfo {title} {Dielectric metasurfaces for complete control of phase and polarization with subwavelength spatial resolution and high transmission},\ }\href {https://doi.org/10.1038/nnano.2015.186} {\bibfield  {journal} {\bibinfo  {journal} {Nature Nanotechnology}\ }\textbf {\bibinfo {volume} {10}}, {\bibinfo {issue} {11}}\ (\bibinfo {year} {2015})}\BibitemShut {NoStop}%
\bibitem [{\citenamefont {Zhang}\ \emph {et~al.}(2023)\citenamefont {Zhang}, \citenamefont {Chang}, \citenamefont {Chen}, \citenamefont {Ding}, \citenamefont {Rahman}, \citenamefont {Duan}, \citenamefont {Stephen},\ and\ \citenamefont {Ni}}]{Zhang2023}%
  \BibitemOpen
  \bibfield  {author} {\bibinfo {author} {\bibfnamefont {L.}~\bibnamefont {Zhang}}, \bibinfo {author} {\bibfnamefont {S.}~\bibnamefont {Chang}}, \bibinfo {author} {\bibfnamefont {X.}~\bibnamefont {Chen}}, \bibinfo {author} {\bibfnamefont {Y.}~\bibnamefont {Ding}}, \bibinfo {author} {\bibfnamefont {M.~T.}\ \bibnamefont {Rahman}}, \bibinfo {author} {\bibfnamefont {Y.}~\bibnamefont {Duan}}, \bibinfo {author} {\bibfnamefont {M.}~\bibnamefont {Stephen}},\ and\ \bibinfo {author} {\bibfnamefont {X.}~\bibnamefont {Ni}},\ }\bibfield  {title} {\bibinfo {title} {High-{{Efficiency}}, 80 mm {{Aperture Metalens Telescope}}},\ }\href {https://doi.org/10.1021/acs.nanolett.2c03561} {\bibfield  {journal} {\bibinfo  {journal} {Nano Letters}\ }\textbf {\bibinfo {volume} {23}},\ \bibinfo {issue} {1} (\bibinfo {year} {2023})}\BibitemShut {NoStop}%
\bibitem [{\citenamefont {Park}\ \emph {et~al.}(2024)\citenamefont {Park}, \citenamefont {Lim}, \citenamefont {Amirzhan}, \citenamefont {Kang}, \citenamefont {Karrfalt}, \citenamefont {Kim}, \citenamefont {Leger}, \citenamefont {Urbas}, \citenamefont {Ossiander}, \citenamefont {Li},\ and\ \citenamefont {Capasso}}]{Park2024}%
  \BibitemOpen
  \bibfield  {author} {\bibinfo {author} {\bibfnamefont {J.-S.}\ \bibnamefont {Park}}, \bibinfo {author} {\bibfnamefont {S.~W.~D.}\ \bibnamefont {Lim}}, \bibinfo {author} {\bibfnamefont {A.}~\bibnamefont {Amirzhan}}, \bibinfo {author} {\bibfnamefont {H.}~\bibnamefont {Kang}}, \bibinfo {author} {\bibfnamefont {K.}~\bibnamefont {Karrfalt}}, \bibinfo {author} {\bibfnamefont {D.}~\bibnamefont {Kim}}, \bibinfo {author} {\bibfnamefont {J.}~\bibnamefont {Leger}}, \bibinfo {author} {\bibfnamefont {A.}~\bibnamefont {Urbas}}, \bibinfo {author} {\bibfnamefont {M.}~\bibnamefont {Ossiander}}, \bibinfo {author} {\bibfnamefont {Z.}~\bibnamefont {Li}},\ and\ \bibinfo {author} {\bibfnamefont {F.}~\bibnamefont {Capasso}},\ }\bibfield  {title} {\bibinfo {title} {All-{{Glass}} 100 mm {{Diameter Visible Metalens}} for {{Imaging}} the {{Cosmos}}},\ }\href {https://doi.org/10.1021/acsnano.3c09462} {\bibfield  {journal} {\bibinfo  {journal} {ACS Nano}\ }\textbf {\bibinfo {volume} {18}},\ \bibinfo {issue} {4} (\bibinfo {year}
  {2024})}\BibitemShut {NoStop}%
\bibitem [{\citenamefont {Majumder}\ \emph {et~al.}(2025)\citenamefont {Majumder}, \citenamefont {Meem}, \citenamefont {Ingold}, \citenamefont {Ricketts}, \citenamefont {Obray}, \citenamefont {Brimhall},\ and\ \citenamefont {Menon}}]{Majumder2025}%
  \BibitemOpen
  \bibfield  {author} {\bibinfo {author} {\bibfnamefont {A.}~\bibnamefont {Majumder}}, \bibinfo {author} {\bibfnamefont {M.}~\bibnamefont {Meem}}, \bibinfo {author} {\bibfnamefont {A.}~\bibnamefont {Ingold}}, \bibinfo {author} {\bibfnamefont {P.}~\bibnamefont {Ricketts}}, \bibinfo {author} {\bibfnamefont {T.}~\bibnamefont {Obray}}, \bibinfo {author} {\bibfnamefont {N.}~\bibnamefont {Brimhall}},\ and\ \bibinfo {author} {\bibfnamefont {R.}~\bibnamefont {Menon}},\ }\bibfield  {title} {\bibinfo {title} {Color astrophotography with a 100 mm-diameter f/2 polymer flat lens},\ }\href {https://doi.org/10.1063/5.0242208} {\bibfield  {journal} {\bibinfo  {journal} {Applied Physics Letters}\ }\textbf {\bibinfo {volume} {126}},\ \bibinfo {issue} {5} (\bibinfo {year} {2025})}\BibitemShut {NoStop}%
\bibitem [{\citenamefont {Rubin}\ \emph {et~al.}(2019)\citenamefont {Rubin}, \citenamefont {D'Aversa}, \citenamefont {Chevalier}, \citenamefont {Shi}, \citenamefont {Chen},\ and\ \citenamefont {Capasso}}]{Rubin2019}%
  \BibitemOpen
  \bibfield  {author} {\bibinfo {author} {\bibfnamefont {N.~A.}\ \bibnamefont {Rubin}}, \bibinfo {author} {\bibfnamefont {G.}~\bibnamefont {D'Aversa}}, \bibinfo {author} {\bibfnamefont {P.}~\bibnamefont {Chevalier}}, \bibinfo {author} {\bibfnamefont {Z.}~\bibnamefont {Shi}}, \bibinfo {author} {\bibfnamefont {W.~T.}\ \bibnamefont {Chen}},\ and\ \bibinfo {author} {\bibfnamefont {F.}~\bibnamefont {Capasso}},\ }\bibfield  {title} {\bibinfo {title} {Matrix {{Fourier}} optics enables a compact full-{{Stokes}} polarization camera},\ }\href {https://doi.org/10.1126/science.aax1839} {\bibfield  {journal} {\bibinfo  {journal} {Science}\ }\textbf {\bibinfo {volume} {365}},\ \bibinfo {pages} {eaax1839} (\bibinfo {year} {2019})}\BibitemShut {NoStop}%
\bibitem [{\citenamefont {Rubin}\ \emph {et~al.}(2022)\citenamefont {Rubin}, \citenamefont {Chevalier}, \citenamefont {Juhl}, \citenamefont {Tamagnone}, \citenamefont {Chipman},\ and\ \citenamefont {Capasso}}]{Rubin2022}%
  \BibitemOpen
  \bibfield  {author} {\bibinfo {author} {\bibfnamefont {N.~A.}\ \bibnamefont {Rubin}}, \bibinfo {author} {\bibfnamefont {P.}~\bibnamefont {Chevalier}}, \bibinfo {author} {\bibfnamefont {M.}~\bibnamefont {Juhl}}, \bibinfo {author} {\bibfnamefont {M.}~\bibnamefont {Tamagnone}}, \bibinfo {author} {\bibfnamefont {R.}~\bibnamefont {Chipman}},\ and\ \bibinfo {author} {\bibfnamefont {F.}~\bibnamefont {Capasso}},\ }\bibfield  {title} {\bibinfo {title} {Imaging polarimetry through metasurface polarization gratings},\ }\href {https://doi.org/10.1364/oe.450941} {\bibfield  {journal} {\bibinfo  {journal} {Optics Express}\ }\textbf {\bibinfo {volume} {30}},\ {\bibinfo {issue} {6}} (\bibinfo {year} {2022})}\BibitemShut {NoStop}%
\bibitem [{\citenamefont {Arbabi}\ \emph {et~al.}(2018)\citenamefont {Arbabi}, \citenamefont {Kamali}, \citenamefont {Arbabi},\ and\ \citenamefont {Faraon}}]{Arbabi2018}%
  \BibitemOpen
  \bibfield  {author} {\bibinfo {author} {\bibfnamefont {E.}~\bibnamefont {Arbabi}}, \bibinfo {author} {\bibfnamefont {S.~M.}\ \bibnamefont {Kamali}}, \bibinfo {author} {\bibfnamefont {A.}~\bibnamefont {Arbabi}},\ and\ \bibinfo {author} {\bibfnamefont {A.}~\bibnamefont {Faraon}},\ }\bibfield  {title} {\bibinfo {title} {Full-{{Stokes Imaging Polarimetry Using Dielectric Metasurfaces}}},\ }\href {https://doi.org/10.1021/acsphotonics.8b00362} {\bibfield  {journal} {\bibinfo  {journal} {ACS Photonics}\ }\textbf {\bibinfo {volume} {5}},\ \bibinfo {issue} {8} (\bibinfo {year} {2018})}\BibitemShut {NoStop}%
\bibitem [{\citenamefont {Trippe}(2014)}]{Trippe2014}%
  \BibitemOpen
  \bibfield  {author} {\bibinfo {author} {\bibfnamefont {S.}~\bibnamefont {Trippe}},\ }\href {https://doi.org/10.5303/JKAS.2014.47.1.015} {\bibinfo {title} {Polarization and {{Polarimetry}}: {{A Review}}}} (\bibinfo {year} {2014}),\ \Eprint {https://arxiv.org/abs/1401.1911} {arXiv:1401.1911 [astro-ph]} \BibitemShut {NoStop}%
\bibitem [{\citenamefont {Kolokolova}(2015)}]{Kolokolova2015}%
  \BibitemOpen
  \bibfield  {author} {\bibinfo {author} {\bibfnamefont {L.}~\bibnamefont {Kolokolova}},\ }\href@noop {} {\emph {\bibinfo {title} {Polarimetry of Stars and Planetary Systems}}}\ (\bibinfo  {publisher} {Cambridge University Press},\ \bibinfo {year} {2015})\BibitemShut {NoStop}%
\bibitem [{\citenamefont {Kolokolova}\ \emph {et~al.}(2015)\citenamefont {Kolokolova}, \citenamefont {Das}, \citenamefont {Dubovik}, \citenamefont {Lapyonok},\ and\ \citenamefont {Yang}}]{Kolokolova2015a}%
  \BibitemOpen
  \bibfield  {author} {\bibinfo {author} {\bibfnamefont {L.}~\bibnamefont {Kolokolova}}, \bibinfo {author} {\bibfnamefont {H.~S.}\ \bibnamefont {Das}}, \bibinfo {author} {\bibfnamefont {O.}~\bibnamefont {Dubovik}}, \bibinfo {author} {\bibfnamefont {T.}~\bibnamefont {Lapyonok}},\ and\ \bibinfo {author} {\bibfnamefont {P.}~\bibnamefont {Yang}},\ }\bibfield  {title} {\bibinfo {title} {Polarization of cosmic dust simulated with the rough spheroid model},\ }\href {https://doi.org/10.1016/j.pss.2015.03.006} {\bibfield  {journal} {\bibinfo  {journal} {Planetary and Space Science}\ }\textbf {\bibinfo {volume} {116}},\ \bibinfo {pages} {30--38} (\bibinfo {year} {2015})}\BibitemShut {NoStop}%
\bibitem [{\citenamefont {Kolokolova}\ \emph {et~al.}(1997)\citenamefont {Kolokolova}, \citenamefont {Jockers}, \citenamefont {Chernova},\ and\ \citenamefont {Kiselev}}]{Kolokolova1997}%
  \BibitemOpen
  \bibfield  {author} {\bibinfo {author} {\bibfnamefont {L.}~\bibnamefont {Kolokolova}}, \bibinfo {author} {\bibfnamefont {K.}~\bibnamefont {Jockers}}, \bibinfo {author} {\bibfnamefont {G.}~\bibnamefont {Chernova}},\ and\ \bibinfo {author} {\bibfnamefont {N.}~\bibnamefont {Kiselev}},\ }\bibfield  {title} {\bibinfo {title} {Properties of cometary dust from color and polarization},\ }\href {https://doi.org/10.1006/icar.1996.5660} {\bibfield  {journal} {\bibinfo  {journal} {Icarus}\ }\textbf {\bibinfo {volume} {126}},\ \bibinfo {issue} {2} (\bibinfo {year} {1997})}\BibitemShut {NoStop}%
\bibitem [{\citenamefont {Rossi}\ and\ \citenamefont {Stam}(2017)}]{Rossi2017}%
  \BibitemOpen
  \bibfield  {author} {\bibinfo {author} {\bibfnamefont {L.}~\bibnamefont {Rossi}}\ and\ \bibinfo {author} {\bibfnamefont {D.~M.}\ \bibnamefont {Stam}},\ }\bibfield  {title} {\bibinfo {title} {Using polarimetry to retrieve the cloud coverage of {{Earth-like}} exoplanets},\ }\href {https://doi.org/10.1051/0004-6361/201730586} {\bibfield  {journal} {\bibinfo  {journal} {Astronomy \& Astrophysics}\ }\textbf {\bibinfo {volume} {607}},\ \bibinfo {pages} {A57} (\bibinfo {year} {2017})}\BibitemShut {NoStop}%
\bibitem [{\citenamefont {Sparks}\ \emph {et~al.}(2021)\citenamefont {Sparks}, \citenamefont {Parenteau}, \citenamefont {Blankenship}, \citenamefont {Germer}, \citenamefont {Patty}, \citenamefont {Bott}, \citenamefont {Telesco},\ and\ \citenamefont {Meadows}}]{Sparks2021}%
  \BibitemOpen
  \bibfield  {author} {\bibinfo {author} {\bibfnamefont {W.~B.}\ \bibnamefont {Sparks}}, \bibinfo {author} {\bibfnamefont {M.~N.}\ \bibnamefont {Parenteau}}, \bibinfo {author} {\bibfnamefont {R.~E.}\ \bibnamefont {Blankenship}}, \bibinfo {author} {\bibfnamefont {T.~A.}\ \bibnamefont {Germer}}, \bibinfo {author} {\bibfnamefont {C.~H.~L.}\ \bibnamefont {Patty}}, \bibinfo {author} {\bibfnamefont {K.~M.}\ \bibnamefont {Bott}}, \bibinfo {author} {\bibfnamefont {C.~M.}\ \bibnamefont {Telesco}},\ and\ \bibinfo {author} {\bibfnamefont {V.~S.}\ \bibnamefont {Meadows}},\ }\bibfield  {title} {\bibinfo {title} {Spectropolarimetry of {{Primitive Phototrophs}} as {{Global Surface Biosignatures}}},\ }\href {https://doi.org/10.1089/ast.2020.2272} {\bibfield  {journal} {\bibinfo  {journal} {Astrobiology}\ }\textbf {\bibinfo {volume} {21}},\ \bibinfo {issue} {2} (\bibinfo {year} {2021})}\BibitemShut {NoStop}%
\bibitem [{\citenamefont {Ellery~Hale}(1979)}]{Hale1979}%
  \BibitemOpen
  \bibfield  {author} {\bibinfo {author} {\bibfnamefont {G.}~\bibnamefont {Ellery~Hale}},\ }\bibfield  {title} {\bibinfo {title} {On the probable existence of a magnetic field in sun-spots},\ }in\ \href@noop {} {\emph {\bibinfo {booktitle} {A Source Book in Astronomy and Astrophysics, 1900--1975}}}\ (\bibinfo  {publisher} {Harvard University Press},\ \bibinfo {year} {1979})\ pp.\ \bibinfo {pages} {96--105}\BibitemShut {NoStop}%
\bibitem [{\citenamefont {Iglesias}\ and\ \citenamefont {Feller}(2019)}]{Iglesias2019}%
  \BibitemOpen
  \bibfield  {author} {\bibinfo {author} {\bibfnamefont {F.~A.}\ \bibnamefont {Iglesias}}\ and\ \bibinfo {author} {\bibfnamefont {A.}~\bibnamefont {Feller}},\ }\bibfield  {title} {\bibinfo {title} {Instrumentation for solar spectropolarimetry: State of the art and prospects},\ }\href@noop {} {\bibfield  {journal} {\bibinfo  {journal} {Optical Engineering}\ }\textbf {\bibinfo {volume} {58}},\ \bibinfo {issue} {8} (\bibinfo {year} {2019})}\BibitemShut {NoStop}%
\bibitem [{\citenamefont {Kosugi}\ \emph {et~al.}(2008)\citenamefont {Kosugi}, \citenamefont {Matsuzaki}, \citenamefont {Sakao}, \citenamefont {Shimizu}, \citenamefont {Sone}, \citenamefont {Tachikawa}, \citenamefont {Hashimoto}, \citenamefont {Minesugi}, \citenamefont {Ohnishi}, \citenamefont {Yamada} \emph {et~al.}}]{Kosugi2008}%
  \BibitemOpen
  \bibfield  {author} {\bibinfo {author} {\bibfnamefont {T.}~\bibnamefont {Kosugi}}, \bibinfo {author} {\bibfnamefont {K.}~\bibnamefont {Matsuzaki}}, \bibinfo {author} {\bibfnamefont {T.}~\bibnamefont {Sakao}}, \bibinfo {author} {\bibfnamefont {T.}~\bibnamefont {Shimizu}}, \bibinfo {author} {\bibfnamefont {Y.}~\bibnamefont {Sone}}, \bibinfo {author} {\bibfnamefont {S.}~\bibnamefont {Tachikawa}}, \bibinfo {author} {\bibfnamefont {T.}~\bibnamefont {Hashimoto}}, \bibinfo {author} {\bibfnamefont {K.}~\bibnamefont {Minesugi}}, \bibinfo {author} {\bibfnamefont {A.}~\bibnamefont {Ohnishi}}, \bibinfo {author} {\bibfnamefont {T.}~\bibnamefont {Yamada}}, \emph {et~al.},\ }\bibfield  {title} {\bibinfo {title} {The {{HINODE}} ({{Solar-B}}) mission: An overview},\ }\href@noop {} {\bibfield  {journal} {\bibinfo  {journal} {Solar Physics} \textbf {\bibinfo {volume} {243}},\ \bibinfo {pages} {3--17}} (\bibinfo {year} {2008})}\BibitemShut {NoStop}%
\bibitem [{\citenamefont {Tsuneta}\ \emph {et~al.}(2008)\citenamefont {Tsuneta}, \citenamefont {Ichimoto}, \citenamefont {Katsukawa}, \citenamefont {Nagata}, \citenamefont {Otsubo}, \citenamefont {Shimizu}, \citenamefont {Suematsu}, \citenamefont {Nakagiri}, \citenamefont {Noguchi}, \citenamefont {Tarbell} \emph {et~al.}}]{Tsuneta2008}%
  \BibitemOpen
  \bibfield  {author} {\bibinfo {author} {\bibfnamefont {S.}~\bibnamefont {Tsuneta}}, \bibinfo {author} {\bibfnamefont {K.}~\bibnamefont {Ichimoto}}, \bibinfo {author} {\bibfnamefont {Y.}~\bibnamefont {Katsukawa}}, \bibinfo {author} {\bibfnamefont {S.}~\bibnamefont {Nagata}}, \bibinfo {author} {\bibfnamefont {M.}~\bibnamefont {Otsubo}}, \bibinfo {author} {\bibfnamefont {T.}~\bibnamefont {Shimizu}}, \bibinfo {author} {\bibfnamefont {Y.}~\bibnamefont {Suematsu}}, \bibinfo {author} {\bibfnamefont {M.}~\bibnamefont {Nakagiri}}, \bibinfo {author} {\bibfnamefont {M.}~\bibnamefont {Noguchi}}, \bibinfo {author} {\bibfnamefont {T.}~\bibnamefont {Tarbell}}, \emph {et~al.},\ }\bibfield  {title} {\bibinfo {title} {The {{Solar Optical Telescope}} for the {{HINODE Mission}}: {{An Overview}}},\ }\href@noop {} {\bibfield  {journal} {\bibinfo  {journal} {Solar Physics}\ }\textbf {\bibinfo {volume} {249}},\ \bibinfo {pages} {167--196} (\bibinfo {year} {2008})}\BibitemShut {NoStop}%
\bibitem [{\citenamefont {West}\ \emph {et~al.}(2011)\citenamefont {West}, \citenamefont {Cirtain}, \citenamefont {Kobayashi}, \citenamefont {Davis}, \citenamefont {Gary},\ and\ \citenamefont {Adams}}]{West2011}%
  \BibitemOpen
  \bibfield  {author} {\bibinfo {author} {\bibfnamefont {E.}~\bibnamefont {West}}, \bibinfo {author} {\bibfnamefont {J.}~\bibnamefont {Cirtain}}, \bibinfo {author} {\bibfnamefont {K.}~\bibnamefont {Kobayashi}}, \bibinfo {author} {\bibfnamefont {J.}~\bibnamefont {Davis}}, \bibinfo {author} {\bibfnamefont {A.}~\bibnamefont {Gary}},\ and\ \bibinfo {author} {\bibfnamefont {M.}~\bibnamefont {Adams}},\ }\bibfield  {title} {\bibinfo {title} {Mg II observations using the MSFC solar ultraviolet magnetograph},\ }in\ \href@noop {} {\emph {\bibinfo {booktitle} {Solar Physics and Space Weather Instrumentation IV}}},\ Vol.\ \bibinfo {volume} {8148}\ (\bibinfo {organization} {SPIE},\ \bibinfo {year} {2011})\ pp.\ \bibinfo {pages} {176--187}\BibitemShut {NoStop}%
\bibitem [{\citenamefont {Zeuner}\ \emph {et~al.}(2022)\citenamefont {Zeuner}, \citenamefont {Belluzzi}, \citenamefont {Guerreiro}, \citenamefont {Ramelli},\ and\ \citenamefont {Bianda}}]{Zeuner2022}%
  \BibitemOpen
  \bibfield  {author} {\bibinfo {author} {\bibfnamefont {F.}~\bibnamefont {Zeuner}}, \bibinfo {author} {\bibfnamefont {L.}~\bibnamefont {Belluzzi}}, \bibinfo {author} {\bibfnamefont {N.}~\bibnamefont {Guerreiro}}, \bibinfo {author} {\bibfnamefont {R.}~\bibnamefont {Ramelli}},\ and\ \bibinfo {author} {\bibfnamefont {M.}~\bibnamefont {Bianda}},\ }\bibfield  {title} {\bibinfo {title} {Hanle rotation signatures in {{Sr I}} 4607 {{{\AA}}}},\ }\href@noop {} {\bibfield  {journal} {\bibinfo  {journal} {Astronomy \& Astrophysics}\ }\textbf {\bibinfo {volume} {662}},\ \bibinfo {pages} {A46} (\bibinfo {year} {2022})}\BibitemShut {NoStop}%
\bibitem [{\citenamefont {Rubin}(2022)}]{Rubin_Polarization_Grating_Design_2023}%
  \BibitemOpen
  \bibfield  {author} {\bibinfo {author} {\bibfnamefont {N.}~\bibnamefont {Rubin}}, } and P. Chevalier, \href@noop {} {\bibinfo {title} {Polarization grating design}} (\bibinfo {year} {2022}), https://github.com/noahrbn/Polarization-Grating-Design\BibitemShut {NoStop}%
\bibitem [{\citenamefont {Li}\ \emph {et~al.}(2023)\citenamefont {Li}, \citenamefont {Rubin}, \citenamefont {Juhl}, \citenamefont {Park},\ and\ \citenamefont {Capasso}}]{Li2023}%
  \BibitemOpen
  \bibfield  {author} {\bibinfo {author} {\bibfnamefont {L.~W.}\ \bibnamefont {Li}}, \bibinfo {author} {\bibfnamefont {N.~A.}\ \bibnamefont {Rubin}}, \bibinfo {author} {\bibfnamefont {M.}~\bibnamefont {Juhl}}, \bibinfo {author} {\bibfnamefont {J.-S.}\ \bibnamefont {Park}},\ and\ \bibinfo {author} {\bibfnamefont {F.}~\bibnamefont {Capasso}},\ }\bibfield  {title} {\bibinfo {title} {Evaluation and characterization of imaging polarimetry through metasurface polarization gratings},\ }\href {https://doi.org/10.1364/AO.480487} {\bibfield  {journal} {\bibinfo  {journal} {Appl. Opt.}\ }\textbf {\bibinfo {volume} {62}},\ \bibinfo {issue} {7} (\bibinfo {year} {2023})}\BibitemShut {NoStop}%
\bibitem [{\citenamefont {{Socas-Navarro}}\ \emph {et~al.}(2006)\citenamefont {{Socas-Navarro}}, \citenamefont {Elmore}, \citenamefont {Pietarila}, \citenamefont {Darnell}, \citenamefont {Lites}, \citenamefont {Tomczyk},\ and\ \citenamefont {Hegwer}}]{Socas2006}%
  \BibitemOpen
  \bibfield  {author} {\bibinfo {author} {\bibfnamefont {H.}~\bibnamefont {{Socas-Navarro}}}, \bibinfo {author} {\bibfnamefont {D.}~\bibnamefont {Elmore}}, \bibinfo {author} {\bibfnamefont {A.}~\bibnamefont {Pietarila}}, \bibinfo {author} {\bibfnamefont {A.}~\bibnamefont {Darnell}}, \bibinfo {author} {\bibfnamefont {B.~W.}\ \bibnamefont {Lites}}, \bibinfo {author} {\bibfnamefont {S.}~\bibnamefont {Tomczyk}},\ and\ \bibinfo {author} {\bibfnamefont {S.}~\bibnamefont {Hegwer}},\ }\bibfield  {title} {\bibinfo {title} {{{SPINOR}}: {{Visible}} and {{Infrared Spectro-Polarimetry}} at the {{National Solar Observatory}}},\ }\href@noop {} {\bibfield  {journal} {\bibinfo  {journal} {Solar Physics}\ }\textbf {\bibinfo {volume} {235}},\ \bibinfo {pages} {55--73} (\bibinfo {year} {2006})}\BibitemShut {NoStop}%
\bibitem [{\citenamefont {Skumanich}\ \emph {et~al.}(1997)\citenamefont {Skumanich}, \citenamefont {Lites}, \citenamefont {Pillet},\ and\ \citenamefont {Seagraves}}]{Skumanich1997}%
  \BibitemOpen
  \bibfield  {author} {\bibinfo {author} {\bibfnamefont {A.}~\bibnamefont {Skumanich}}, \bibinfo {author} {\bibfnamefont {{\relax BW}.}~\bibnamefont {Lites}}, \bibinfo {author} {\bibfnamefont {V.~M.}\ \bibnamefont {Pillet}},\ and\ \bibinfo {author} {\bibfnamefont {P.}~\bibnamefont {Seagraves}},\ }\bibfield  {title} {\bibinfo {title} {The {{Calibration}} of the {{Advanced Stokes Polarimeter}}},\ }\href@noop {} {\bibfield  {journal} {\bibinfo  {journal} {The Astrophysical Journal Supplement Series}\ }\textbf {\bibinfo {volume} {110}},\ \bibinfo {issue} {2} (\bibinfo {year} {1997})}\BibitemShut {NoStop}%
\bibitem [{\citenamefont {{Socas-Navarro}}\ \emph {et~al.}(2011)\citenamefont {{Socas-Navarro}}, \citenamefont {Elmore}, \citenamefont {Ramos},\ and\ \citenamefont {Harrington}}]{Socas-Navarro2011}%
  \BibitemOpen
  \bibfield  {author} {\bibinfo {author} {\bibfnamefont {H.}~\bibnamefont {{Socas-Navarro}}}, \bibinfo {author} {\bibfnamefont {D.}~\bibnamefont {Elmore}}, \bibinfo {author} {\bibfnamefont {A.~A.}\ \bibnamefont {Ramos}},\ and\ \bibinfo {author} {\bibfnamefont {{\relax DM}.}~\bibnamefont {Harrington}},\ }\bibfield  {title} {\bibinfo {title} {Characterization of telescope polarization properties across the visible and near-infrared spectrum-{{Case}} study: The {{Dunn Solar Telescope}}},\ }\href@noop {} {\bibfield  {journal} {\bibinfo  {journal} {Astronomy \& Astrophysics}\ }\textbf {\bibinfo {volume} {531}},\ \bibinfo {pages} {A2} (\bibinfo {year} {2011})}\BibitemShut {NoStop}%
\bibitem [{\citenamefont {Lu}\ and\ \citenamefont {Chipman}(1996)}]{Lu1996}%
  \BibitemOpen
  \bibfield  {author} {\bibinfo {author} {\bibfnamefont {S.-Y.}\ \bibnamefont {Lu}}\ and\ \bibinfo {author} {\bibfnamefont {R.~A.}\ \bibnamefont {Chipman}},\ }\bibfield  {title} {\bibinfo {title} {Interpretation of {{Mueller}} matrices based on polar decomposition},\ }\href {https://doi.org/10.1364/JOSAA.13.001106} {\bibfield  {journal} {\bibinfo  {journal} {J. Opt. Soc. Am. A}\ }\textbf {\bibinfo {volume} {13}},\ \bibinfo {issue} {5} (\bibinfo {year} {1996})}\BibitemShut {NoStop}%
\bibitem [{\citenamefont {Degl'Innocenti}\ and\ \citenamefont {Landolfi}(2006)}]{Degl2006}%
  \BibitemOpen
  \bibfield  {author} {\bibinfo {author} {\bibfnamefont {M.~L.}\ \bibnamefont {Degl'Innocenti}}\ and\ \bibinfo {author} {\bibfnamefont {M.}~\bibnamefont {Landolfi}},\ }\href@noop {} {\emph {\bibinfo {title} {Polarization in Spectral Lines}}}\ (\bibinfo  {publisher} {Springer Science \& Business Media},\ \bibinfo {year} {2006})\BibitemShut {NoStop}%
\bibitem [{\citenamefont {Jovanovic}\ \emph {et~al.}(2023)\citenamefont {Jovanovic}, \citenamefont {Gatkine}, \citenamefont {Anugu}, \citenamefont {{Amezcua-Correa}}, \citenamefont {Thakur}, \citenamefont {Beichman}, \citenamefont {Bender}, \citenamefont {Berger}, \citenamefont {Bigioli}, \citenamefont {{Bland-Hawthorn}} \emph {et~al.}}]{Jovanovic2023}%
  \BibitemOpen
  \bibfield  {author} {\bibinfo {author} {\bibfnamefont {N.}~\bibnamefont {Jovanovic}}, \bibinfo {author} {\bibfnamefont {P.}~\bibnamefont {Gatkine}}, \bibinfo {author} {\bibfnamefont {N.}~\bibnamefont {Anugu}}, \bibinfo {author} {\bibfnamefont {R.}~\bibnamefont {{Amezcua-Correa}}}, \bibinfo {author} {\bibfnamefont {R.~B.}\ \bibnamefont {Thakur}}, \bibinfo {author} {\bibfnamefont {C.}~\bibnamefont {Beichman}}, \bibinfo {author} {\bibfnamefont {C.~F.}\ \bibnamefont {Bender}}, \bibinfo {author} {\bibfnamefont {J.-P.}\ \bibnamefont {Berger}}, \bibinfo {author} {\bibfnamefont {A.}~\bibnamefont {Bigioli}}, \bibinfo {author} {\bibfnamefont {J.}~\bibnamefont {{Bland-Hawthorn}}}, \emph {et~al.},\ }\bibfield  {title} {\bibinfo {title} {2023 {{Astrophotonics Roadmap}}: {{Pathways}} to {{Realizing Multi-Functional Integrated Astrophotonic Instruments}}},\ }\href@noop {} {\bibfield  {journal} {\bibinfo  {journal} {Journal of Physics: Photonics}\ }\textbf {\bibinfo {volume} {5}},\ \bibinfo {issue} {4} (\bibinfo
  {year} {2023})}\BibitemShut {NoStop}%
\bibitem [{\citenamefont {Cvetojevic}\ \emph {et~al.}(2012)\citenamefont {Cvetojevic}, \citenamefont {Jovanovic}, \citenamefont {Lawrence}, \citenamefont {Withford},\ and\ \citenamefont {{Bland-Hawthorn}}}]{Cvetojevic2012}%
  \BibitemOpen
  \bibfield  {author} {\bibinfo {author} {\bibfnamefont {N.}~\bibnamefont {Cvetojevic}}, \bibinfo {author} {\bibfnamefont {N.}~\bibnamefont {Jovanovic}}, \bibinfo {author} {\bibfnamefont {J.}~\bibnamefont {Lawrence}}, \bibinfo {author} {\bibfnamefont {M.}~\bibnamefont {Withford}},\ and\ \bibinfo {author} {\bibfnamefont {J.}~\bibnamefont {{Bland-Hawthorn}}},\ }\bibfield  {title} {\bibinfo {title} {Developing arrayed waveguide grating spectrographs for multi-object astronomical spectroscopy},\ }\href@noop {} {\bibfield  {journal} {\bibinfo  {journal} {Optics Express}\ }\textbf {\bibinfo {volume} {20}},\ \bibinfo {issue} {3} (\bibinfo {year} {2012})}\BibitemShut {NoStop}%
\bibitem [{\citenamefont {Gatkine}\ \emph {et~al.}(2017)\citenamefont {Gatkine}, \citenamefont {Veilleux}, \citenamefont {Hu}, \citenamefont {{Bland-Hawthorn}},\ and\ \citenamefont {Dagenais}}]{Gatkine2017}%
  \BibitemOpen
  \bibfield  {author} {\bibinfo {author} {\bibfnamefont {P.}~\bibnamefont {Gatkine}}, \bibinfo {author} {\bibfnamefont {S.}~\bibnamefont {Veilleux}}, \bibinfo {author} {\bibfnamefont {Y.}~\bibnamefont {Hu}}, \bibinfo {author} {\bibfnamefont {J.}~\bibnamefont {{Bland-Hawthorn}}},\ and\ \bibinfo {author} {\bibfnamefont {M.}~\bibnamefont {Dagenais}},\ }\bibfield  {title} {\bibinfo {title} {Arrayed {{Waveguide Grating Spectrometers}} for {{Astronomical Applications}}: {{New Results}}},\ }\href@noop {} {\bibfield  {journal} {\bibinfo  {journal} {Optics express}\ }\textbf {\bibinfo {volume} {25}},\ \bibinfo {issue} {15} (\bibinfo {year} {2017})}\BibitemShut {NoStop}%
\bibitem [{\citenamefont {{Bland-Hawthorn}}\ and\ \citenamefont {Kern}(2012)}]{Bland2012}%
  \BibitemOpen
  \bibfield  {author} {\bibinfo {author} {\bibfnamefont {J.}~\bibnamefont {{Bland-Hawthorn}}}\ and\ \bibinfo {author} {\bibfnamefont {P.}~\bibnamefont {Kern}},\ }\bibfield  {title} {\bibinfo {title} {Molding the {{Flow}} of {{Light}}: {{Photonics}} in {{Astronomy}}},\ }\href@noop {} {\bibfield  {journal} {\bibinfo  {journal} {Physics Today}\ }\textbf {\bibinfo {volume} {65}},\ \bibinfo {issue} {5} (\bibinfo {year} {2012})}\BibitemShut {NoStop}%
\bibitem [{\citenamefont {{Leon-Saval}}\ \emph {et~al.}(2013)\citenamefont {{Leon-Saval}}, \citenamefont {Argyros},\ and\ \citenamefont {{Bland-Hawthorn}}}]{Leon2013}%
  \BibitemOpen
  \bibfield  {author} {\bibinfo {author} {\bibfnamefont {S.~G.}\ \bibnamefont {{Leon-Saval}}}, \bibinfo {author} {\bibfnamefont {A.}~\bibnamefont {Argyros}},\ and\ \bibinfo {author} {\bibfnamefont {J.}~\bibnamefont {{Bland-Hawthorn}}},\ }\bibfield  {title} {\bibinfo {title} {Photonic {{Lanterns}}},\ }\href@noop {} {\bibfield  {journal} {\bibinfo  {journal} {Nanophotonics}\ }\textbf {\bibinfo {volume} {2}},\ \bibinfo {pages} {5--6} (\bibinfo {year} {2013})}\BibitemShut {NoStop}%
\bibitem [{\citenamefont {Moraitis}\ \emph {et~al.}(2021)\citenamefont {Moraitis}, \citenamefont {{Alvarado-Zacarias}}, \citenamefont {{Amezcua-Correa}}, \citenamefont {Jeram},\ and\ \citenamefont {Eikenberry}}]{Moraitis2021}%
  \BibitemOpen
  \bibfield  {author} {\bibinfo {author} {\bibfnamefont {C.~D.}\ \bibnamefont {Moraitis}}, \bibinfo {author} {\bibfnamefont {J.~C.}\ \bibnamefont {{Alvarado-Zacarias}}}, \bibinfo {author} {\bibfnamefont {R.}~\bibnamefont {{Amezcua-Correa}}}, \bibinfo {author} {\bibfnamefont {S.}~\bibnamefont {Jeram}},\ and\ \bibinfo {author} {\bibfnamefont {S.~S.}\ \bibnamefont {Eikenberry}},\ }\bibfield  {title} {\bibinfo {title} {Demonstration of high-efficiency photonic lantern couplers for {{PolyOculus}}},\ }\href@noop {} {\bibfield  {journal} {\bibinfo  {journal} {Applied Optics}\ }\textbf {\bibinfo {volume} {60}},\ \bibinfo {issue} {19} (\bibinfo {year} {2021})}\BibitemShut {NoStop}%
\bibitem [{\citenamefont {De~Wijn}\ \emph {et~al.}(2022)\citenamefont {De~Wijn}, \citenamefont {Casini}, \citenamefont {Carlile}, \citenamefont {Lecinski}, \citenamefont {Sewell}, \citenamefont {Zmarzly}, \citenamefont {Eigenbrot}, \citenamefont {Beck}, \citenamefont {W{\"o}ger},\ and\ \citenamefont {Kn{\"o}lker}}]{DeWijn2022}%
  \BibitemOpen
  \bibfield  {author} {\bibinfo {author} {\bibfnamefont {A.~G.}\ \bibnamefont {De~Wijn}}, \bibinfo {author} {\bibfnamefont {R.}~\bibnamefont {Casini}}, \bibinfo {author} {\bibfnamefont {A.}~\bibnamefont {Carlile}}, \bibinfo {author} {\bibfnamefont {{\relax AR}.}~\bibnamefont {Lecinski}}, \bibinfo {author} {\bibfnamefont {S.}~\bibnamefont {Sewell}}, \bibinfo {author} {\bibfnamefont {P.}~\bibnamefont {Zmarzly}}, \bibinfo {author} {\bibfnamefont {{\relax AD}.}~\bibnamefont {Eigenbrot}}, \bibinfo {author} {\bibfnamefont {C.}~\bibnamefont {Beck}}, \bibinfo {author} {\bibfnamefont {F.}~\bibnamefont {W{\"o}ger}},\ and\ \bibinfo {author} {\bibfnamefont {M.}~\bibnamefont {Kn{\"o}lker}},\ }\bibfield  {title} {\bibinfo {title} {The {{Visible Spectro-Polarimeter}} of the {{Daniel K}}. {{Inouye Solar Telescope}}},\ }\href@noop {} {\bibfield  {journal} {\bibinfo  {journal} {Solar Physics}\ }\textbf {\bibinfo {volume} {297}},\ \bibinfo {issue} {22} (\bibinfo {year} {2022})}\BibitemShut {NoStop}%
\end{thebibliography}


\end{document}